\documentclass[12pt]{article}
\usepackage{graphicx}
\usepackage{titlesec}
\usepackage[hyperfootnotes=true]{hyperref}
\hypersetup{colorlinks, citecolor=blue, linkcolor=blue, urlcolor=blue}
\usepackage{makecell}

\usepackage{hyperref}%
\usepackage{color,soul}
\usepackage{amsmath}
\usepackage{makecell}
\usepackage{multirow} 
\usepackage{booktabs}
\usepackage{svg}
\svgpath{{./figs}} %
\providecommand{\keywords}[1]
{\small	
\textbf{\textit{Keywords---}} #1}
\usepackage{geometry}
\graphicspath{{figs/}}
\usepackage{caption}
\usepackage{listings}
\usepackage{amssymb}
\usepackage{pifont}
\usepackage{caption}
\usepackage[round]{natbib}
\let\cite\citep % Redefine \cite to act like \citep
\usepackage{eqns}  % Use the equations
\usepackage{authblk}

\title{\textbf{Challenges and Advancements in Modeling Shock Fronts with Physics-Informed Neural Networks: A Review and Benchmarking Study}}

\author[1]{Jassem Abbasi\thanks{Corresponding author: Jassem Abbasi (jassem.abbasi@uis.no)}}
\author[2]{Ameya D. Jagtap}
\author[3]{Ben Moseley}
\author[1]{Aksel Hiorth}
\author[1]{Pål Østebø Andersen}

\affil[1]{\textit{Department of Energy Resources, University of Stavanger, Stavanger, Norway}}
\affil[2]{\textit{Aerospace Engineering Department, Worcester Polytechnic Institute, Worcester, MA 01609, USA}}
\affil[3]{\textit{Department of Earth Science and Engineering, Imperial College London, London, UK}}
\date{}

\begin{document}
\maketitle
\begin{abstract}
Solving partial differential equations (PDEs) with discontinuous solutions — such as shock waves in multiphase viscous flow in porous media — is critical for a wide range of scientific and engineering applications, as they represent sudden changes in physical quantities. Physics-Informed Neural Networks (PINNs), an approach proposed for solving PDEs, encounter significant challenges when applied to such systems. Accurately solving PDEs with discontinuities using PINNs requires specialized techniques to ensure effective solution accuracy and numerical stability. Various methods have been developed to address the challenges of modeling discontinuities within the PINNs framework. This work reviews and benchmarks these approaches across problems of varying complexity, categorizing them into three broad groups,  influencing solution accuracy differently. (1) Physics-modification (PM) methods improve accuracy by modifying the system's physics, such as adding artificial viscosity or enforcing entropy constraints. (2) Loss and training modification (LM) techniques focus on regularizing the loss landscape, often by refining the loss term in high-error regions. (3) Architecture-modification (AM) approaches, on the other hand, propose advanced network designs to handle discontinuities better.
A benchmarking study was conducted on two multiphase flow problems in porous media: the classic Buckley-Leverett (BL) problem and a fully coupled system of equations involving shock waves but with varying levels of solution complexity. The findings show that PM and LM approaches can provide accurate solutions for the BL problem by effectively addressing the infinite gradients associated with shock occurrences. In contrast, AM methods failed to effectively resolve the shock waves. When applied to fully coupled PDEs (with more complex loss landscape), the generalization error in the solutions quickly increased, highlighting the need for ongoing innovation.
This study provides a comprehensive review of existing techniques for managing PDE discontinuities using PINNs, offering information on their strengths and limitations. The results underscore the necessity for further research to improve PINNs ability to handle complex discontinuities, particularly in more challenging problems with complex loss landscapes. This includes problems involving higher dimensions or multiphysics systems, where current methods often struggle to maintain accuracy and efficiency.

\end{abstract}

\keywords{Deep learning, Multiphase flow, Porous media, Physics-informed neural networks (PINNs), Buckley-Leverett, Shock modeling}

% \end{document}

\newpage 

\tableofcontents
\newpage

\section{Introduction}

Discontinuities in solutions of PDEs, characterized by abrupt changes in system properties such as velocity, pressure, or phase volume fractions, are ubiquitous in scientific and engineering applications. These discontinuities often manifest as shock waves, contact discontinuities, or cusps, and they play a critical role in phenomena such as high-speed aerodynamics \cite{Gaitonde2015ProgressInteractions}, astrophysical explosions \cite{Longair2011HighAstrophysics}, molecular transport in materials \cite{Mattila2016ASimulations, Kodama2002ShockCells}, and multiphase flow in porous media \cite{Zhang2024Physics-InformedSolubility, Sahimi2011FlowApproaches}. In the context of fluid flow in porous media, discontinuities arise due to the complex interplay of viscous, capillary, and inertial forces, leading to sudden variations in fluid properties that challenge traditional modeling and simulation approaches \cite{Mitra2020FrontsCapillarity}.

Mathematically, discontinuities in PDE solutions are associated with the breakdown of smoothness in the solution profile, often resulting from nonlinearities in the governing equations. For instance, in multiphase flow in porous media, the governing equations are derived from mass conservation laws and Darcy's law, which relate the volumetric flow rate to fluid viscosity, permeability, and pressure gradients \cite{Bear2013DynamicsMedia}. When viscous forces dominate—such as in high-pressure gradients or coarse-grained porous materials—the system exhibits shock fronts, which are discontinuities in the solution. These shocks are inherently difficult to resolve numerically due to their non-smooth nature and the potential for nonphysical oscillations or numerical diffusion when using conventional discretization methods. Traditional numerical techniques for solving PDEs, such as Finite-Difference (FD), Finite-Element (FE), and Finite-Volume (FV) methods, have been extensively employed to model such systems \cite{Griffiths2006NumericalEngineers}. However, capturing discontinuities accurately requires specialized shock-capturing schemes, such as total variation diminishing (TVD) methods, weighted essentially non-oscillatory (WENO) schemes, or adaptive mesh refinement (AMR). These methods aim to stabilize the numerical solution near discontinuities while maintaining high-order accuracy in smooth regions. Despite significant progress over the past decades, challenges remain in balancing computational efficiency, accuracy, and robustness, particularly for complex multiphysics problems \cite{Rawat2010OnInteractions}.

Recently, Physics-Informed Neural Networks (PINNs) have emerged as a promising alternative for solving PDEs, leveraging the universal approximation capabilities of neural networks to encode physical laws directly into the loss function \cite{Raissi2019Physics-informedEquations}. PINNs offer several advantages, including mesh-free discretization, seamless handling of high-dimensional problems, and the ability to incorporate experimental or observational data. However, vanilla-PINNs struggle to resolve discontinuities effectively, often leading to inaccurate or oscillatory solutions near shocks \cite{Liu2024DiscontinuityNetworks, Wang2024UnderstandingPINNs}. This limitation stems from the smoothness of neural network representations and the difficulty of optimizing the loss landscape in the presence of sharp gradients. In general, \citet{DeRyck2024NumericalLearning} demonstrated the estimated error of physics-informed machine learning methods can be significantly impacted by the stability of the underlying PDEs (PDEs change with respect to perturbations). To address these challenges, several strategies have been proposed, including adaptive weighting of loss terms \cite{Liu2021ShockWaves}, domain decomposition \cite{Jagtap2020ExtendedEquations}, and the use of artificial viscosity \cite{Fuks2020LIMITATIONSMEDIA}. Each of these methods offers unique advantages for resolving discontinuities. For instance, adaptive loss weighting prioritizes regions near shocks, domain decomposition isolates discontinuities for specialized treatment, and artificial viscosity introduces controlled diffusion to stabilize solutions near sharp gradients. Although these advances have shown promise, a complete understanding of their strengths and limitations remains incomplete, particularly for complex problems such as multiphase flow in porous media, where the interplay of physical forces complicates the resolution of discontinuity \cite{Abbasi2024History-MatchingPINNs}.

In this study, we comprehensively review the various strategies for addressing PDE discontinuities using PINNs. Then, we benchmark these strategies on two problems: a coupled two-phase flow problem and a simplified formulation of the same problem, both of which feature varying levels of convexity in the loss landscape. By evaluating recent advances in shock-capturing techniques for PINNs, we aim to identify practical methods that enhance their applicability to problems involving shock waves. Our analysis provides insights into the trade-offs between accuracy, computational cost, and robustness, offering guidance for future research in this area.

The remainder of this paper is organized as follows. Section 2 provides an overview of discontinuities and shock waves in PDEs, highlighting their mathematical and physical characteristics. Section 3 discusses the governing equations for multiphase flow in porous media, emphasizing the role of viscous and capillary forces. Section 4 introduces the concept of PINNs and their application to PDEs. Section 5 reviews existing shock-capturing strategies for PINNs, while Section 6 presents a comparative benchmarking of these strategies across two example problems. Finally, Section 7 concludes with a discussion of key findings and future research directions.

\begin{figure}[h]
    \centering
    \includegraphics[width=17cm]{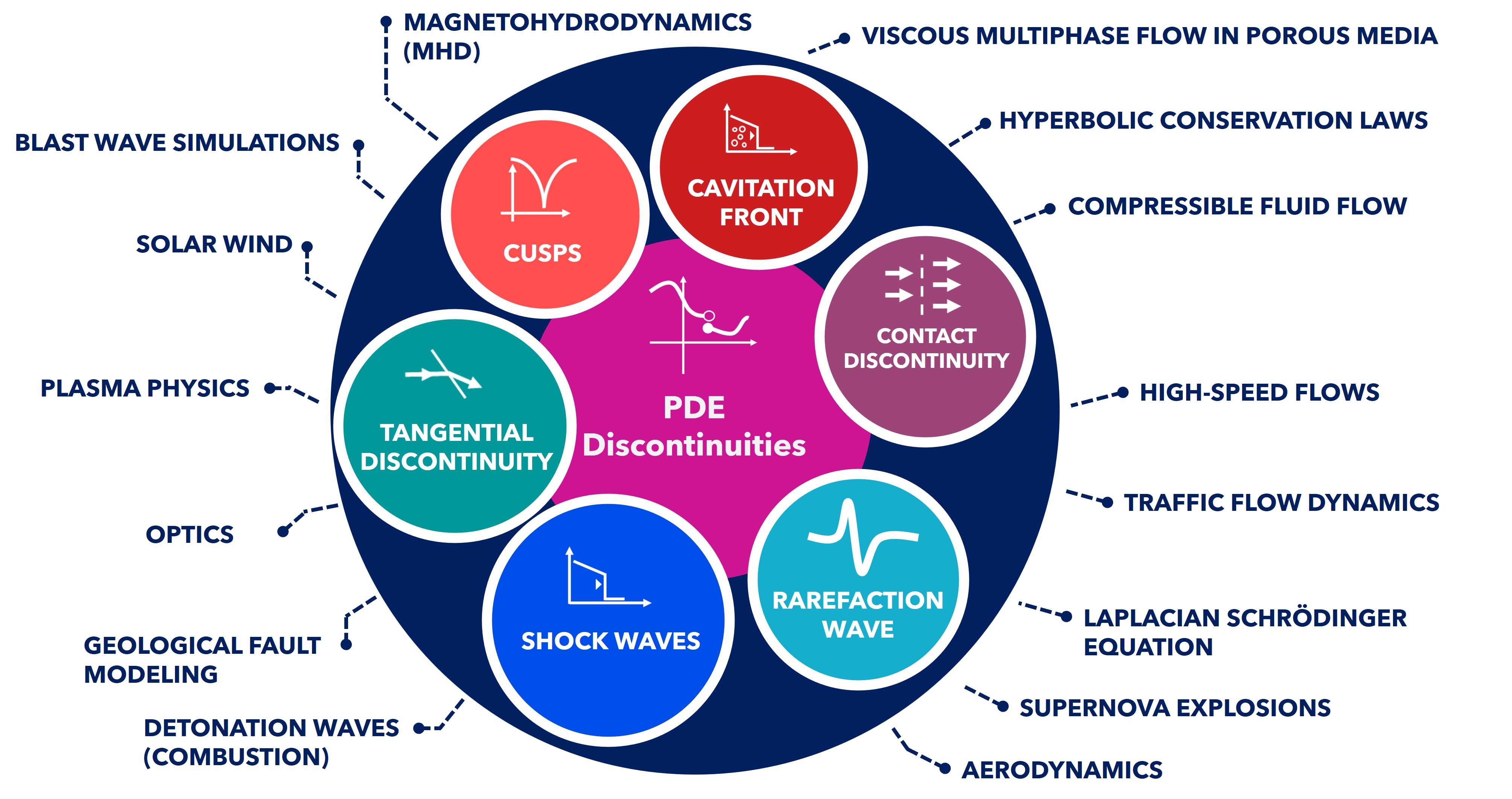}
    \caption{Various types of PDE discontinuities, including jumps and kinks with real-world physical examples. }
    \label{fig:introduction}
\end{figure}

% {\color{red}Jassem! We do encounter jump discontinuities not only in high-speed flows governed by hyperbolic conservation laws (HCL) but also in MHD, shallow water, and traffic flow problems. I think we should make this sketch more general.}

\section{Discontinuities in PDEs: An Overview}

Discontinuities in PDEs occur when the solution or its derivatives exhibit abrupt changes (jumps) across specific regions or interfaces in the domain \cite{LeVeque1992NumericalLaws}. These discontinuities are inherent to many physical processes and pose significant challenges for both analytical and numerical solutions of PDEs. Figure \ref{fig:introduction} provides an overview of the major types of discontinuities in PDEs and their real-world implications, while Table~\ref{table:discont_summary} summarizes their key characteristics. Discontinuities can be classified into two broad categories: strong discontinuities, where the solution itself exhibits abrupt jumps, and weak discontinuities, where the solution remains continuous, but its derivatives experience abrupt changes \cite{LANDAU198714Diffusion}. Additionally, quasi-discontinuities involve smooth yet rapid variations in system variables over a finite region, which, while not true discontinuities, still present significant computational challenges. Moreover, arbitrary or initial-state discontinuities, which often arise as initial/boundary conditions (unrelated to conservation laws), can complicate the analysis and solution of PDEs.

\begin{figure}
    \centering
    \includegraphics[width=0.5\linewidth]{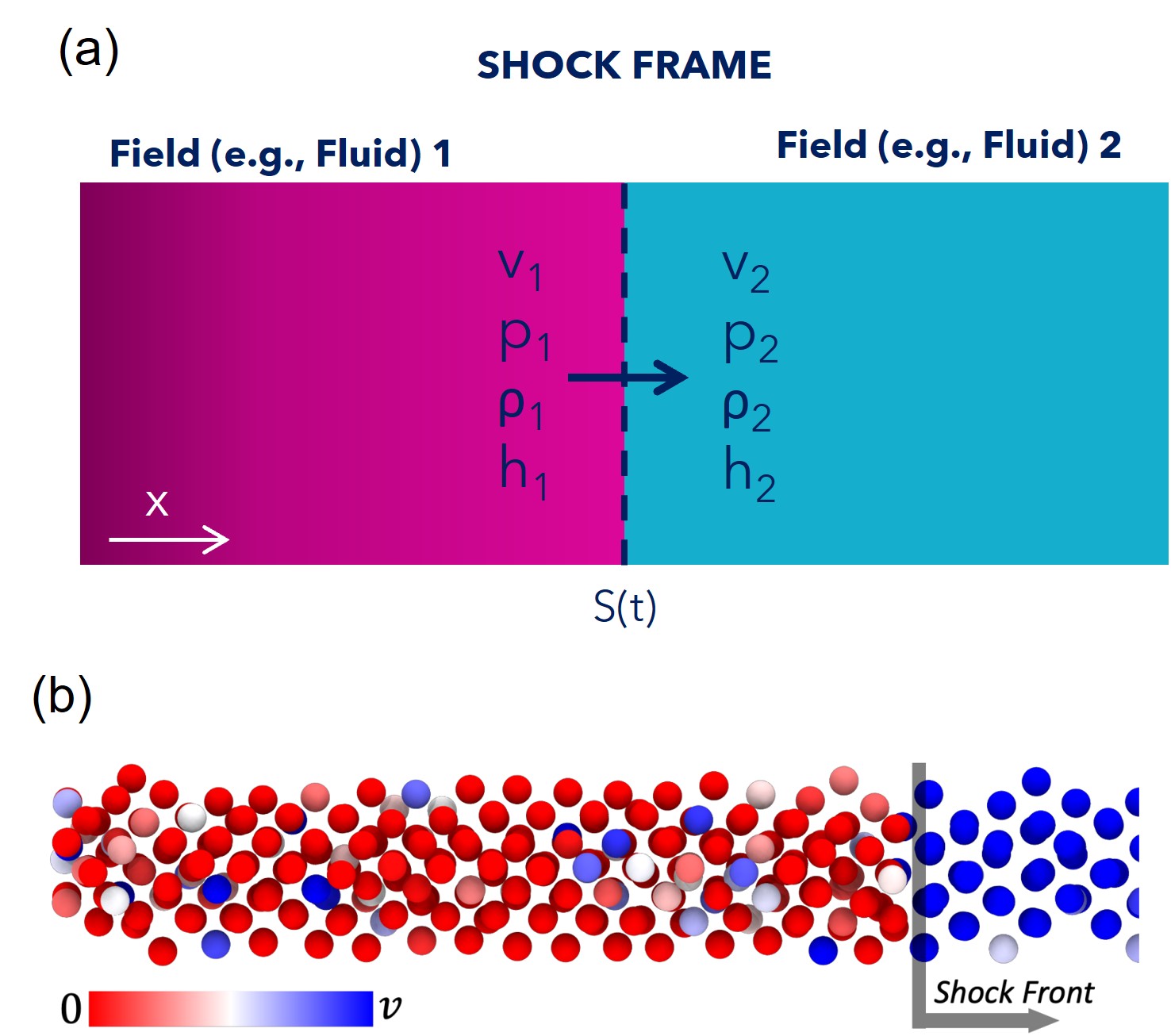}
    \caption{ A simplified visualization of a 1D shock-wave frame at the interface of two fields (e.g., flowing fluids), a) a continuum-scale perspective, b) a molecular-scale perspective \cite{Patel2022ThermodynamicallySystems}.}
    \label{fig:RankieHugoniot}
\end{figure}

Shock waves are the most prominent type of strong discontinuities, characterized by abrupt jumps in field properties (e.g., density, velocity, and pressure) across an interface. They commonly arise in hyperbolic PDEs that govern conservation laws in fluid dynamics \cite{LeVeque1992NumericalLaws}. To analyze the behavior across a shock, a reference frame fixed to the discontinuity surface is often used, with the x-axis oriented normal to the surface (see Fig. \ref{fig:RankieHugoniot}). For inviscid flows, the conservation laws across the discontinuity are expressed by the Rankine-Hugoniot conditions:

\begin{equation}
\rho_1 v_{1} = \rho_2 v_{2},
\label{eq:massbalance}
\end{equation}

\begin{equation}
\rho_1 v_{1}^2 + p_1 = \rho_2 v_{2}^2 + p_2, \quad 
\label{eq:momentumbalance}
\end{equation}

\begin{equation}
 \frac{1}{2}v_1^2 + h_1 = \frac{1}{2}v_2^2 + h_2.
\label{eq:energybalance}
\end{equation}

The equations above represent the conservation of mass, momentum, and energy, respectively. Here, subscripts \(1\) and \(2\) indicate the states on either side of the shock discontinuity, \(\rho\) represents the fluid density, \(v_x\) is the velocity component normal to the discontinuity, and \(h\) denotes the specific enthalpy of the fluid.

For shock waves, in a stationary reference frame, mass flux across the discontinuity is nonzero, i.e., \(p_1 v_1 = p_2 v_2 \neq 0\). Here, variables such as density, velocity, and pressure can change abruptly due to non-linear wave interactions. These interactions render the system inherently non-linear and result in an irreversible increase in entropy \cite{Maltby2023LocalWaves}. Mathematically, a shock wave can be represented as a moving surface \(S(t)\) in space-time across which the solution \(u(x,t)\) experiences a discontinuous jump. In cavitation fronts, shocks can induce phase changes, increasing the susceptibility of computations to errors \cite{Ezzatneshan2020SimulationState}. In shock waves, despite the discontinuous nature of the physical solution, it satisfies the integral form of conservation laws, leading to what are termed \textbf{weak solutions} \cite{MishraNumericalEquations}.

In addition to shock waves, there may be other types of strong discontinuities including \textbf{contact discontinuities}, \textbf{tangential discontinuities}, and \textbf{weak discontinuities}. For instance, tangential discontinuities occur where the field properties change abruptly without any normal mass flux across the interface (\(p_1 v_1 = p_2 v_2 = 0\)) \cite{Huang2010FLUIDDISCONTINUITY}. Weak discontinuities are a special type of discontinuity, such as cusps, where the function remains continuous, but its derivatives exhibit abrupt changes. These phenomena, commonly observed in solutions to mechanical problems (such as fracture propagation in solid materials), introduce singularities in mathematical modeling and result in a loss of accuracy in numerical computations \cite{Moffatt2019SingularitiesMechanics}.

Each type of discontinuity — e.g., shock waves, tangential discontinuities, and weak discontinuities — introduces distinct difficulties in the analysis and computation of PDEs, such as numerical instabilities, conservation errors, and increased computational cost during the solution process \cite{LeVeque1992NumericalLaws}.

In addition to the aforementioned discontinuities, problems involving quasi-discontinuities - such as rarefaction and deflagration combustion waves (Table \ref{table:discont_summary}) — present challenges in computational fluid dynamics. Resolving quasi-discontinuities numerically is particularly demanding due to the need for high spatial and temporal resolution to accurately capture the steep gradients without introducing excessive numerical dissipation or dispersion. Advanced computational techniques, such as adaptive mesh refinement (AMR), high-order schemes, and shock-capturing methods, are often employed to address these challenges effectively. For an overview of weak discontinuities and their properties, refer to Table \ref{table:discont_summary}.

\begin{table}[h!]
\centering
\caption{Classification of Discontinuities and Singularities in Hyperbolic PDEs: Strong discontinuities involve abrupt jumps ($[\cdot] \neq 0$) in flow variables, and weak discontinuities involve smooth variations or weaker jumps. Here, $\mathbf{u}$ is velocity, $\mathbf{n}$ is the normal vector, and $p$, $\rho$, $s$ are pressure, density, and entropy, respectively.}
\label{table:discont_summary}
\small % Reduce font size for the table
\begin{tabular}{llll}
\toprule
\textbf{Type} & \textbf{Pressure} & \textbf{Velocity} & \textbf{Density/Entropy} \\
\midrule
\multicolumn{4}{l}{\textbf{Discontinuity (Abrupt Change)}} \\
\midrule
Shock Wave & $[p] \neq 0$ & $[\mathbf{u}] \neq 0$ & $[\rho] \neq 0$, $[s] \neq 0$ \\
Contact Discontinuity & $[p] = 0$ & $[\mathbf{u} \cdot \mathbf{n}] = 0$ & $[\rho] \neq 0$, $[s] \neq 0$ \\
Tang. Discontinuity & $[p] = 0$ & $\mathbf{u} \cdot \mathbf{n} = 0$, $[\mathbf{u}_\text{tang}] \neq 0$ & $[\rho] \neq 0$, $[s] \neq 0$ \\
Slip Line & $[p] = 0$ & $[\mathbf{u} \cdot \mathbf{n}] = 0$, $[\mathbf{u}_\text{tang}] \neq 0$ & $[\rho] \neq 0$, $[s] \neq 0$ \\
Current Sheet & $[p] \neq 0$ & $[\mathbf{u}] \neq 0$ & $[\rho] \neq 0$, $[s] \neq 0$ \\
Cavitation Front & $[p] \neq 0$ & $[\mathbf{u}] \neq 0$ & $[\rho] \neq 0$, $[s] \neq 0$ + phase change\\
Weak Discontinuity (Cusp) & $[\nabla p] \neq 0 $  & $[\nabla u] \neq 0 $ & $[\nabla\rho] \neq 0 $, $[\nabla s] \neq 0$ \\
\midrule
\multicolumn{4}{l}{\textbf{Quasi-Discontinuity (Smooth Change)}} \\
\midrule
Rarefaction Wave & $[p]\neq 0$  & $\mathbf{[u]\neq 0}$  & $[\rho]\neq 0$, $[s]= 0$ \\
Deflagration Wave & $[p] \neq 0$ & $[\mathbf{u}] \neq 0$ & $[\rho] \neq 0$, $[s] \neq 0$ \\
\bottomrule
\end{tabular}
\end{table}

In numerical analysis, various mathematical techniques have been developed to address these challenges in the last decades, including weak formulation, entropy conditions, flux limiters, upwind schemes, artificial viscosity, and high/adaptive resolution schemes \cite{GodunovFiniteDynamics, Pirozzoli2011NumericalFlows}. In theory, the thickness of the shock wave is usually on the order of a few mean free paths of the fluid molecules, resulting in an extremely thin layer compared to the overall flow field \cite{Kundu2012FluidMechanics}. There is a general interest in resolving the sharpest possible shock profiles while avoiding nonphysical oscillations. However, these two objectives often conflict, as reducing the shock width can introduce numerical instabilities, while excessive smoothing diminishes the accuracy of the solution. There is a consensus that there is no single "correct" shape for a numerical solution of the mathematically discontinuous shock profile; rather, the representation of shocks varies based on the numerical method employed, as discussed by \citet{Margolin2022ArtificialNow}. For instance, artificial viscosity methods often trade some accuracy in favor of smoother profiles, effectively damping sharp gradients \cite{Carpenter1998ComputationalFlows}. High-resolution schemes, on the other hand, minimize numerical artifacts and better capture sharp transitions, though they may result in reduced differentiability at the shock \cite{Toro2009RiemannIntroduction}. The analysis reveals that in higher spatial dimensions, complexity rises significantly due to multidimensional phenomena such as shock curvature and complex wave interactions \cite{Hu2024AFlows, Fleischmann2020ADissipation}. 

In this work, we review the various methodologies proposed for capturing shock fronts using PINNs and provide a benchmark comparison of the most prominent approaches, based on the analysis of cases of multiphase flow in porous media (however, the conclusions can be applied to other shock-front phenomena). In the context of multiphase flow in porous media, flooding of one phase by another phase can generate a shock wave, which leads to an abrupt change of phase properties. These discontinuities can have a critical effect on the accuracy of flow predictions, the precise location of the front, and the reliability of inverse calculations. 

\section{Theory of Multiphase Flow in Porous Media}
Multiphase flow in porous media may be investigated on several scales, such as pore scale and continuum scale. We focus on the continuum-scale in which flow is described by Darcy's law. For phase $i$, the Darcy velocity ($v_i$) is \cite{Muccino1998TowardMedia}:

\begin{equation}
v_i =-\lambda_i (\nabla p_i - \rho_i g),\ \ \lambda_i=\frac{Kk_{ri}}{\mu_i},\ \ \label{eq:darcy}
\end{equation}

where $\lambda_i$ is the mobility of phase $i$ and $p_i$ is the phase pressure. Also, $K$ is the absolute permeability tensor,  $k_{ri}$ is the phase relative permeability and $\rho_i$ the phase density. Moreover, $g$ is the gravitational acceleration vector. For each phase, we define a mass conservation law to constrain the spatiotemporal phase transport: \cite{Bear2013DynamicsMedia},
\begin{equation}
\partial_t\left(\phi\rho_is_i\right)=-\nabla\cdot\left(\rho_i v_i\right),\ \ \ {x\in\Omega} \label{eq:diff2pflow}
\end{equation}

where $\phi$ refers to the rock porosity, and $s_i$ refers to the phase saturation (fluid volume fraction of the pore space). The equation above is a hyperbolic conservation law that relates the unknown state variables $s_i$ and $p_i$ to the independent spatiotemporal variables $t$ and $x$. The PDEs for the different phases are coupled with a constraint that the phase volume fractions add up to 1, i.e. $\sum_{i=1}^{N} s_i = 1$, and that the phases locally have a pressure difference defined by a capillary pressure function $p_c$ \cite{Bear2013DynamicsMedia}. 

For a two-phase system (wetting phase $w$ and non-wetting phase $nw$), the mass conservation equations simplify under the assumptions of incompressible fluids, horizontal flow (neglecting gravity), and negligible capillary pressure ($p_w = p_{nw}$). The resulting system of equations is:

\begin{equation}
\begin{cases}
    \phi  \frac{\partial s_w}{\partial t} - \frac{K }{\mu_w}  \frac{d}{dx} \left( k_{rw} \frac{d p_{nw}}{dx} \right) = 0, \\
    \phi \frac{\partial (1- s_{w})}{\partial t} - \frac{K }{\mu_{nw}} \frac{d}{dx} \left( k_{rnw}  \frac{d p_{nw}}{dx} \right) = 0, \\
\end{cases}
,\ \ \ x \in \Omega
\label{eq:systemof1d2pflow}
\end{equation}

These equations must be solved simultaneously as a coupled system to capture the interdependencies between the phases. The independent variables of the system are $x$ and $t$, and the dependent (state) variables are $s_w$ and $p_{nw}$. The state variables can be determined by setting appropriate initial and boundary conditions. The initial condition refers to the values of saturation and pressure at the starting time, i.e., $t = 0$. These are written as $s_w(x, t=0)$ and $p_{nw}(x, t=0)$. For a system under constant pressure, the boundary conditions are given by: $s_w(x=0, t) = s_{w,inlet}$, $p(x=0, t) = p_{inlet}$, and $p(x=L, t) = p_{outlet}$.

Under constant injection rate conditions, the Buckley-Leverett equation further simplifies the system by eliminating the pressure variable and introducing the fractional flow function $f_w(s_w)$. The resulting hyperbolic equation is:

\begin{equation}
\ensuremath{  \frac{\partial {s_{w}}}{\partial t} + \frac{v_t}{\phi}\frac{d {f_{w}}}{d s_w}\frac{\partial {s_{w}}}{\partial x} = 0 ,\ \ 
\left(i=w,nw\right) }
\label{eq:BLeq}
\end{equation}

where $v_t = v_w + v_{nw}$ is the total Darcy velocity, and $f_w(s_w)$ is the fractional flow function of the wetting phase. This equation describes the formation of a shock front followed by a rarefaction wave, characteristic of displacement processes in porous media. More details on these equations and their applications on one-dimensional flooding scenarios under constant pressure drop and constant injection rate (Buckley-Leverett problem) are discussed in Appendices \ref{app:constant_pressure} and \ref{app:buckley_leverett}, respectively.

\begin{figure}[htp]
    \centering
    \includegraphics[width=6cm]{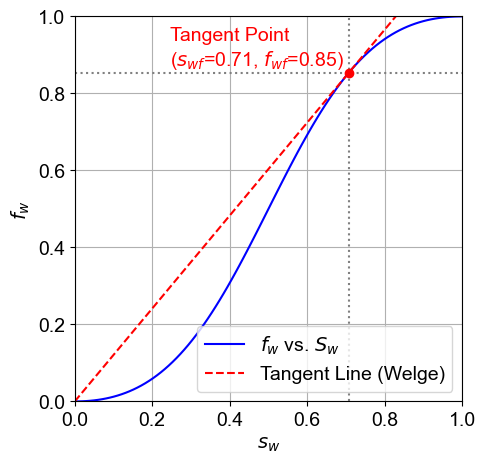}
    \caption{A graphical visualization of fractional flow $f_w$ vs. $s_w$ and tangent line (Welge method). The applied properties are as follows: $k_{ri}^{max}$=1.0, $n_{i}$=2.0, and $\mu_{i}$=1.0 cP. }
    \label{fig:welge}
\end{figure}

\section{Physics-Informed Neural Networks (PINNs)}

As an emerging technique for solving PDEs, PINNs integrate physical principles into the neural network's training process, by adding the differential equations to a composite loss function \cite{Raissi2019Physics-informedEquations}. Although PINNs work for any PDE in general, here consider a physical system described by a hyperbolic PDE $\mathcal{D}$ in the domain \( \Omega \):
\begin{align}
\mathcal{D}(u) := \frac{\partial u}{\partial t} + \nabla \cdot f(u) = 0, \quad \mathbf{x} \in \Omega, \, t \in [0, T],  \label{eq:pdedef}
\end{align}
and with the initial condition
\begin{align}
u = u_0, \quad \mathbf{x} \in \Omega, \, t \in [0],\label{eq:icpdedef}
\end{align}
Then, we define $\mathcal{N}^{\theta}$ as a non-linear neural network that gives the solution of the above system  
\begin{align}
\mathcal{N}^{\theta}(u^{\theta}(\mathbf{x}, t)) , \quad \mathbf{x} \in \Omega, \, t \in [0, T],
\label{eq:pinnsoperator}
\end{align}
where \( u^{\theta}(\mathbf{x}, t) \) is the neural solution of the problem, \( \mathbf{x} \) is the spatial coordinate(s) of the system, and \( t \) represents time in the range \( [0, T]\). 
A neural network is used to approximate the solution \( u(\mathbf{x}, t) \). Let \( u(\mathbf{x}, t) \) be the true solution of the PDE $\mathcal{D}$, then the goal is to train the neural network such that \( u^{\theta}(\mathbf{x}, t) \approx u(\mathbf{x}, t) \). To this end, we define the \textit{generalization error} as the difference between the true solution and the neural solution, 
\begin{align}
\varepsilon_G = \| u^{\theta} - u \|
\label{eq:generalizationerror}
\end{align}
Also, with respect to the investigated PDE and the underlying boundary conditions, we may define the following residuals:
\begin{align}
\mathcal{R}_{\mathcal{D}}(x, t) = \frac{\partial u^{\theta}}{\partial t} + \nabla \cdot f(u^{\theta})
\label{eq:residuals}
\end{align}
\begin{align}
\mathcal{R}_{\text{IC}}(x) = u^{\theta}(x, 0) - u_0(x)
\label{eq:icres}
\end{align}
\begin{align}
\mathcal{R}_{\text{BC}}(y, t) = u^{\theta}(y, t) - g(y, t),
\label{eq:bcres}
\end{align}
where, $\mathcal{R}_{\mathcal{D}}$ represents the residual of the PDE, $\mathcal{R}_{\text{IC}}$ denotes the deviations at the initial conditions (IC), and $\mathcal{R}_{\text{BC}}$ indicates the deviations at the boundary conditions (BC). To train the neural network \(\mathcal{N}^{\theta}\), the residuals defined above are computed at specific collocation points distributed across various spatiotemporal locations within the domain of interest. These points include the initial condition region and the spatial boundaries of the system. Then, it leads to the definition of the average training error as,
\begin{align}
\varepsilon_T := w_{\mathcal{D}} \left( \frac{1}{N} \sum_{n=1}^{N} |\mathcal{R}_{\mathcal{D}}|^p \right)^{\frac{1}{p}}+ w_{tb}\left( \frac{1}{N_{tb}} \sum_{n=1}^{N_{tb}}  |\mathcal{R}_{IC}|^p \right)^{\frac{1}{p}} + w_{sb}\left( \frac{1}{N_{xb}} \sum_{n=1}^{N_{xb}}  |\mathcal{R}_{BC}|^p \right)^{\frac{1}{p}} 
\label{eq:et}
\end{align}
where p is a positive real number (typically $p \ge 1$) that defines the type of norm being applied to the residuals.

In the training process, we define the training error as the loss function, i.e., $\mathcal{L}_T = \varepsilon_T$ in which we are looking for minimizing it, by optimizing the network trainable parameters in a way that,
\begin{align}
\theta = argmin({{\mathcal{L}}}_{T}).
\label{eq:argminfor}
\end{align}

The minimization of the loss function is achieved through iterative tuning of the network parameters $\theta$, typically using gradient-based optimization methods such as SGD, Adam, or their variants. 

Shock fronts present challenges for PINNs due to their discontinuous nature. In problems like two-phase flow in porous media, where sharp interfaces form between phases, abrupt changes in fluid properties (e.g., pressure, saturation) violate the smoothness assumptions inherent in PINNs. The key challenges include:

\begin{itemize}
    \item \textbf{Loss Function Limitations}: 
    The objective functions (e.g., $L^2$-norm), used in PINNs to quantify residuals, are inherently biased toward smooth solutions and struggle to resolve high-frequency or steep gradients. As a result, the loss function tends to penalize extremely sharp changes, leading to overly smooth predictions that fail to capture shock fronts accurately \cite{Krishnapriyan2021CharacterizingNetworks}.
    \item \textbf{Gradient-Based Optimization}: Discontinuities provide non-smooth loss landscapes, which restrict convergence and lead to poor shock front approximation \cite{Wang2020UnderstandingNetworks}. Training of neural networks relies on gradient computations, which require continuity in both the neural network $\mathcal{N}^{\theta}$ and the PDEs $\mathcal{D}$ being modeled. This continuity assumption poses a fundamental limitation for training PINNs, particularly for problems involving discontinuities such as shock fronts \cite{de2024error,DeRyck2022ErrorPDEs}.
    \item \textbf{Spectral Bias}: Neural networks are typically biased toward low-frequency characteristics, making it challenging to adequately describe high-frequency phenomena such as discontinuities (strong or quasi discontinuities) \cite{Rahaman2019OnNetworks,Wang2022WhenPerspective}.
\end{itemize}

To address these issues, techniques such as adaptive loss weighting, shock-capturing schemes, and hybrid approaches combining PINNs with traditional numerical methods have been proposed. In the following section, we explore these methods, focusing on their application to the problem of two-phase flow in porous media.

\section{Shock Capturing Techniques}
\label{section:shock_techniques}
Various approaches have been proposed to address the limitations of PINNs when dealing with discontinuities in the PDEs. Table \ref{table:pinnsmethodshockall} summarizes a range of proposed techniques that aim to enhance the performance of PINNs to capture discontinuities. 
This review primarily focuses on search keywords related to the investigation of key mathematical concepts for improving the performance of PINNs, including \textit{shock waves/fronts}, \textit{PDE discontinuities}, and \textit{stiff PDEs}, within the time range of 2018-2024. The proposed methods are tested across various types of discontinuities in PDEs, including shock fronts in conservation equations (such as multiphase flow in porous media), rarefaction waves (oblique or expansion waves), Burgers' equation, Sod shock tube problems. These techniques aim to enhance the model's ability to resolve discontinuities without sacrificing physical accuracy. In this work, we focus on reviewing and comparing these proposed techniques, with particular attention to their performance in capturing shock fronts in problems of multiphase flow in porous media.

First, we provide a brief overview of the historical development of modeling shocks using PINNs. \citet{Fuks2020LIMITATIONSMEDIA, Almajid2022PredictionNetworks} were among the first to identify the challenges that PINNs encounter in accurately capturing shock fronts in two-phase flow in porous media. Building on established techniques in numerical methods, \citet{Fuks2020LIMITATIONSMEDIA} applied the concept of artificial diffusion to smooth out discontinuities. This strategy laid the groundwork for subsequent improvements. In more recent works, attention mechanisms have been introduced to enhance PINNs performance in regions with sharp gradients. \citet{Rodriguez-Torrado2022Physics-informedProblem} and \citet{Diab2022Data-FreeProblem} applied discretized attention and adaptive mechanisms to focus on critical areas near shocks, helping improve the accuracy of the solution in these challenging regions. Other strategies, such as Integral PINNs \cite{RajvanshiINTEGRALLAWS} and weak formulations \cite{DeRyck2024WPINNs:Laws, Chaumet2024ImprovingSystems}, have been employed to address discontinuities in a broader range of PDEs beyond the Buckley-Leverett problem. These methods leverage the integral form or weak formulations of the governing equations, improving the handling of shock waves, rarefaction waves, and other discontinuous phenomena. More recent innovations include gradient-weighted PINNs \cite{Liu2024DiscontinuityNetworks}, relaxation neural networks \cite{Zhou2024CapturingNetworks}, and extended PINNs (XPINNs) with entropy conditions \cite{Jagtap2022Physics-informedFlows}, which are designed to handle more complex flow behaviors and shock interactions in multidimensional problems. 

Next, we provide a detailed review of the proposed shock modeling methods. We will classify them into three subgroups, 

\begin{enumerate}
    \item \textbf{Physics modification methods (PM),} which modify the solved PDE to remove the shock;
    \item  \textbf{Loss and training modification methods (LM)} which modify the loss landscape of the network during the training and (or) the training algorithm to promote convergence;
    \item   \textbf{Architecture modification methods (AM),}  which modify the network architecture to improve its ability to capture non-linearities in the solution.
    \end{enumerate}

\begin{figure}[h]
    \centering
    \includegraphics[width=0.98\textwidth]{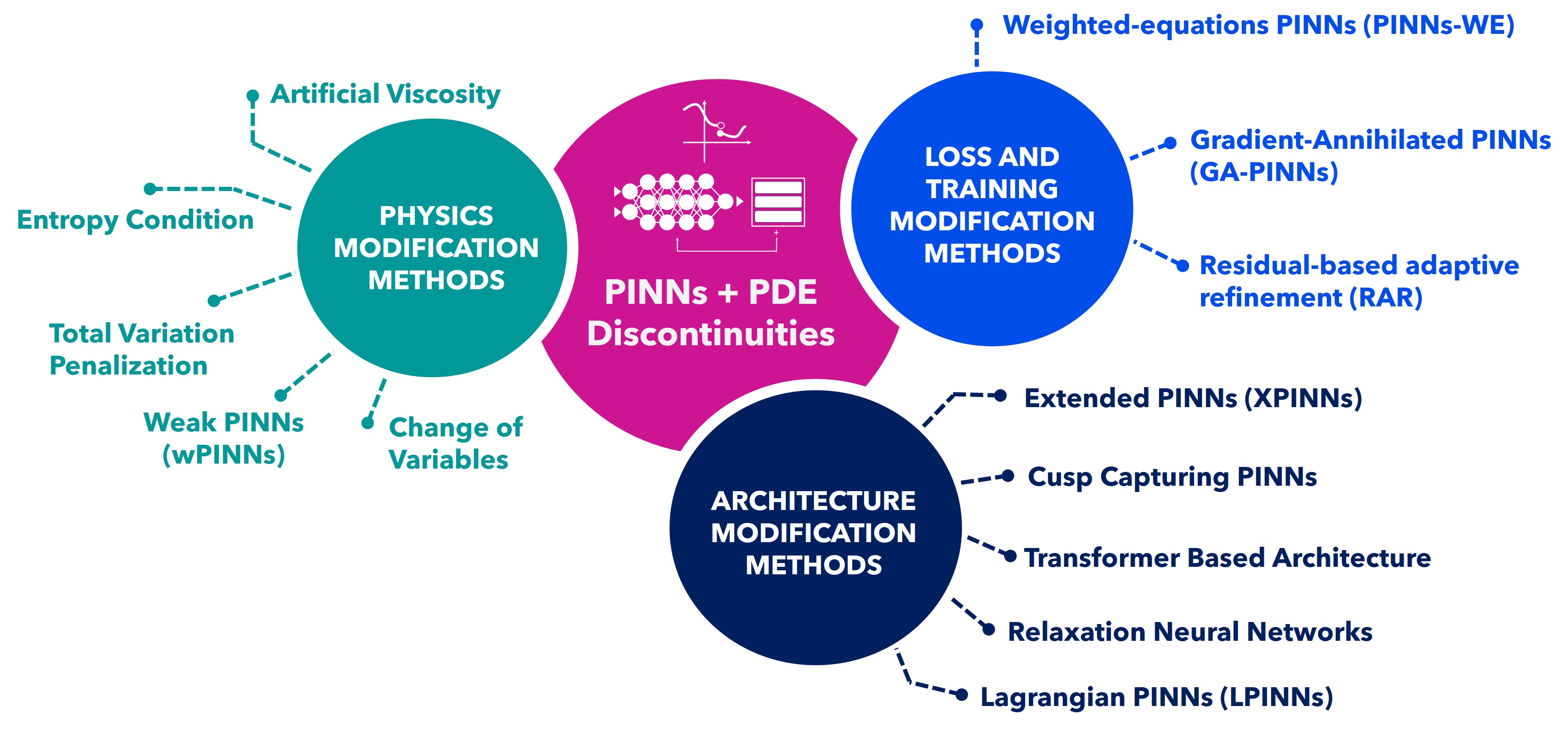}
    \caption{A classification of shock management techniques for PINNs.}
    \label{fig:figclasses}
\end{figure}

These subgroups are visualized in Figure \ref{fig:figclasses}. The final section of this review provides a detailed benchmark comparison of the most promising methods for addressing the shock wave issue in two-phase flow problems in porous media.

\begin{table}[!tp]
\caption{A summary of the studies relevant for improving the performance of PINNs in solving problems with discontinuities. The table also shows which types of shocks have been investigated in each work. Also, the class(es) of the proposed techniques (PM, LM or AM), based on classifications provided in section \ref{section:shock_techniques}, is determined. 
Additionally, the types of investigated discontinuities are clarified as either Discontinuity [D] or Quasi-discontinuity [QD], based on the classifications provided in Table \ref{table:discont_summary}.
}
\begin{center}
\resizebox{\textwidth}{!}{%
    \label{table:pinnsmethodshockall}
    \begin{tabular}{|l|l|l|c|c|}
    \hline
    \multirow{2}{*}{\textbf{Reference}} & \multirow{2}{*}{\textbf{Technique [Class]}} & \multirow{2}{*}{\textbf{PDE(s)}} & \multicolumn{2}{|c|}{\textbf{\makecell[l]{Shock \\ Type}}} \\ 
    &&& \rotatebox{-90}{\textbf{Static}}& \rotatebox{-90}{\textbf{Dynamic}} \\ \hline
    \citet{Fuks2020LIMITATIONSMEDIA} & Artificial Diffusion [PM] & Buckley-Leverett [D] & \ding{55}& \ding{51} \\ \hline
    \citet{mao2020physics} & \makecell[l]{Improved \\collocation points [LM]} & \makecell[l]{1D/2D Euler equations \\ Expansion waves \\ 2D Oblique waves [D, QD]} & \ding{51}& \ding{51} \\ \hline
    \citet{Papados2021SolvingExtension} & \makecell[l]{Weighted-PINNs [LM] \\ Domain Extension} & \makecell[l]{Sod problem [D]} & \ding{51}& \ding{51} \\ \hline
    \citet{Rodriguez-Torrado2022Physics-informedProblem}& Discritized Attention [AM] & Buckley-Leverett [D] & \ding{55}& \ding{51} \\ \hline
    \citet{Diab2022Data-FreeProblem} & Attention Mechanism [AM] & Buckley-Leverett [D]& \ding{55}& \ding{51} \\ \hline

    \citet{Patel2022ThermodynamicallySystems} & \makecell[l]{cvPINNs [PM], \\ Total Variation} & \makecell[l]{Buckley-Leverett, \\ Copper hydrodynamics, \\ Burger's equation, etc. [D]} & \ding{51}& \ding{51} \\ \hline
    \citet{Jagtap2022Physics-informedFlows} & \makecell[l]{XPINNs, \\ Entropy conditions [PM, AM]} & \makecell[l]{Euler equations \\ 2D Expansion waves \\ 2D Oblique waves [D, QD]} & \ding{51}& \ding{55} \\ \hline
    \citet{Laubscher2022ModelingSolver} & \makecell[l]{Fourier features [AM]} & \makecell[l]{Inviscid flow} & \ding{51}& \ding{55} \\ \hline
    \citet{Braga-Neto2022Characteristics-InformedProblems} & \makecell[l]{CINNs [AM]} & \makecell[l]{Advection PDE [D]} & \ding{51}& \ding{51} \\ \hline
    \citet{Coutinho2023Physics-informedViscosity} & Adaptive Viscosity [PM, LM] & \makecell[l]{Buckley-Leverett, \\ Burgers Equation [D]} & \ding{51}& \ding{51} \\ \hline
    \citet{Mojgani2023KolmogorovPDEs}  & \makecell[l]{LPINNs [AM, LM]} & \makecell[l]{Reaction–diffusion, \\ Convection–diffusion \\ Burgers’ equation [D, QD]} & \ding{51}& \ding{55} \\ \hline
    \citet{Ferrer-Sanchez2023GRADIENT-ANNIHILATEDPREPRINT} & \makecell[l]{GA-PINNs [LM]} & \makecell[l]{Riemann problem \\ Sod Shock Tube [D]} & \ding{51}& \ding{51} \\ \hline
    \citet{Tseng2023AForces}  & \makecell[l]{Cusp Capturing PINNs [AM]} & \makecell[l]{2D/3D Stokes [D]} & \ding{51}& \ding{55} \\ \hline
    \citet{Ma2024Physics-InformedEquation}  & \makecell[l]{PIGANs [AM]} & \makecell[l]{Buckley–Leverett [D]} & \ding{51}& \ding{51} \\ \hline
    \citet{RajvanshiINTEGRALLAWS}  & Integral PINNs [LM]& \makecell[l]{Shallow Water, \\ Buckley-Leverett [D]} & \ding{55}& \ding{51} \\ \hline
    \citet{DeRyck2024WPINNs:Laws}  & Weak Formulation [PM]& \makecell[l]{Standing shock, \\ Moving Shock, \\ Rarefaction Wave, etc.[D, QD]} & \ding{51}& \ding{51} \\ \hline
    \citet{Chaumet2024ImprovingSystems} & Weak Formulation [PM] & \makecell[l]{Standing shock, \\ Moving Shock, \\ Rarefaction Wave, etc. [D, QD]} & \ding{51}& \ding{51} \\ \hline
    
    \citet{Liu2024DiscontinuityNetworks}  & \makecell[l]{Gradient-weighted  [LM]\\ PINNs} & \makecell[l]{Burger's equation, \\ 2D Riemann problem, \\ Transonic flow, etc. [D, QD]} & \ding{55}& \ding{51} \\ \hline
    \citet{Zhou2024CapturingNetworks}  & Relaxation NNs [AM]& \makecell[l]{Burger's equation, \\ Sod problem, \\ Shallow water, etc. [D]} & \ding{51}& \ding{51} \\ \hline
    \citet{Zhang2024Physics-InformedSolubility}  & \makecell[l]{Welge’s construction [PM]} & \makecell[l]{Buckley-Leverett [D]} & \ding{55}& \ding{51} \\ \hline
    \citet{Wang2024Chien-physics-informedProblems} & Chien-PINNs [AM] & \makecell[l]{Elliptic PDEs [D, QD]} & \ding{51}& \ding{55} \\ \hline
    \citet{Patsatzis2024GoRINNs:Laws}  & \makecell[l]{GoRINNs [AM, LM]} & \makecell[l]{Traffic Flow \\ Burger's equation [D]} & \ding{51}& \ding{55} \\ \hline
    \end{tabular}
}
\end{center}
\end{table}

\subsection{Physics Modification Methods}
A more scientific approach to addressing the issue of shock waves is to fundamentally reformulate the physical problem, to mitigate the inherent challenge of gradient explosion, which often arises in such scenarios. It helps to reduce the likelihood of abrupt numerical instabilities and improving the overall reliability of the solution, at the potential expense of reducing modeling accuracy.

\subsubsection{Artificial Viscosity} 

The concept of using artificial viscosity to handle shock waves was first introduced by \citet{GodunovFiniteDynamics}, who proposed a finite-difference approach to address discontinuous solutions in fluid flow problems. This technique later became a fundamental tool in Computational Fluid Dynamics (CFD) solvers, evolving in various forms to improve numerical stability and accuracy of the solutions. Artificial viscosity is applied to convert frontal kinetic energy into internal energy, serving as an efficient mechanism to induce the required increase in entropy across shock waves \cite{Albright2020LocallyHydrodynamics}. More recently, it has been demonstrated that artificial viscosity can also enhance the performance of PINNs when dealing with discontinuities. 

One of the earliest applications in this context was by \citet{Fuks2020LIMITATIONSMEDIA}, who applied PINNs to solve the Buckley-Leverett (BL) problem, a one-dimensional hyperbolic PDE. Their results revealed significant limitations in the ability of vanilla-PINNs, which refers to the standard version of PINNs proposed by \citet{Raissi2019Physics-informedEquations}, to accurately solve the problem and replicate fluid fronts. To address these issues, they proposed adding a diffusion term, which smooths the discontinuity at the fronts, thus enabling more accurate solutions. This technique can be written by extending the eq. \ref{eq:BLeq} as, 
\begin{equation}
\frac{\partial s_w}{\partial t} + \frac{\partial f(s_w)}{\partial x} = \epsilon \frac{\partial^2 s_w}{\partial x^2},
\end{equation}

Adding the diffusion term $\epsilon \frac{\partial^2 s_w}{\partial x^2}$ above helps reduce the hyperbolicity of the system, helping the PINNs solver to find the solution of the system more effectively. The coefficient  $\epsilon$, known as the artificial viscosity coefficient, plays a crucial role in controlling the amount of diffusion applied to the system. To maintain the physical accuracy of the simulation, \( \epsilon << 1\), ensuring that while the discontinuities are smoothed, the essential characteristics of the system—such as sharp fronts—are not overly smeared. This balance is key to stabilizing the solution process while preserving the fidelity of the system's dynamics. In numerical methods, $\epsilon$ is chosen to be proportional to the grid spacing $\Delta x$ and the magnitude of the gradient of the solution \cite{Margolin2022ArtificialNow},
\begin{equation}
\epsilon \propto \Delta x \left| \frac{\partial s_i}{\partial x} \right|
\end{equation}
\citet{Fuks2020LIMITATIONSMEDIA} proposed $\epsilon=0.004$, and \citet{Almajid2022PredictionNetworks} proposed  $\epsilon=0.001$. The generality of the values for various system properties has not been discussed.

However, in the case of solving full form PDEs of flow in porous media, i.e., eq. \ref{eq:systemof1d2pflow}, the conservation law is rewritten as:
\begin{equation}
\phi \frac{\partial (\rho_i s_i)}{\partial t} + \frac{\partial (\rho_i v_i)}{\partial x} - \epsilon_i \frac{\partial^2 s_i}{\partial x^2} = 0
\label{eq:artificialviscosity2p}
\end{equation}
where $\epsilon_i$ is separately defined for each phase. As an alternative approach, the Darcy velocity equation was modified to account for the spatial gradient of water saturation, incorporating an additional term (gradient of water saturation) that influences the system's behavior  \cite{Cogswell2017SimulationTimesteps}:
\begin{equation}
\vec{v}_i = -k \lambda_i \left[ \vec{\nabla} \bar{p_i} - \epsilon_i \vec{\nabla} s_i \right]
\end{equation}

\begin{figure}[h!]
    \centering
    \includegraphics[width=0.7\linewidth]{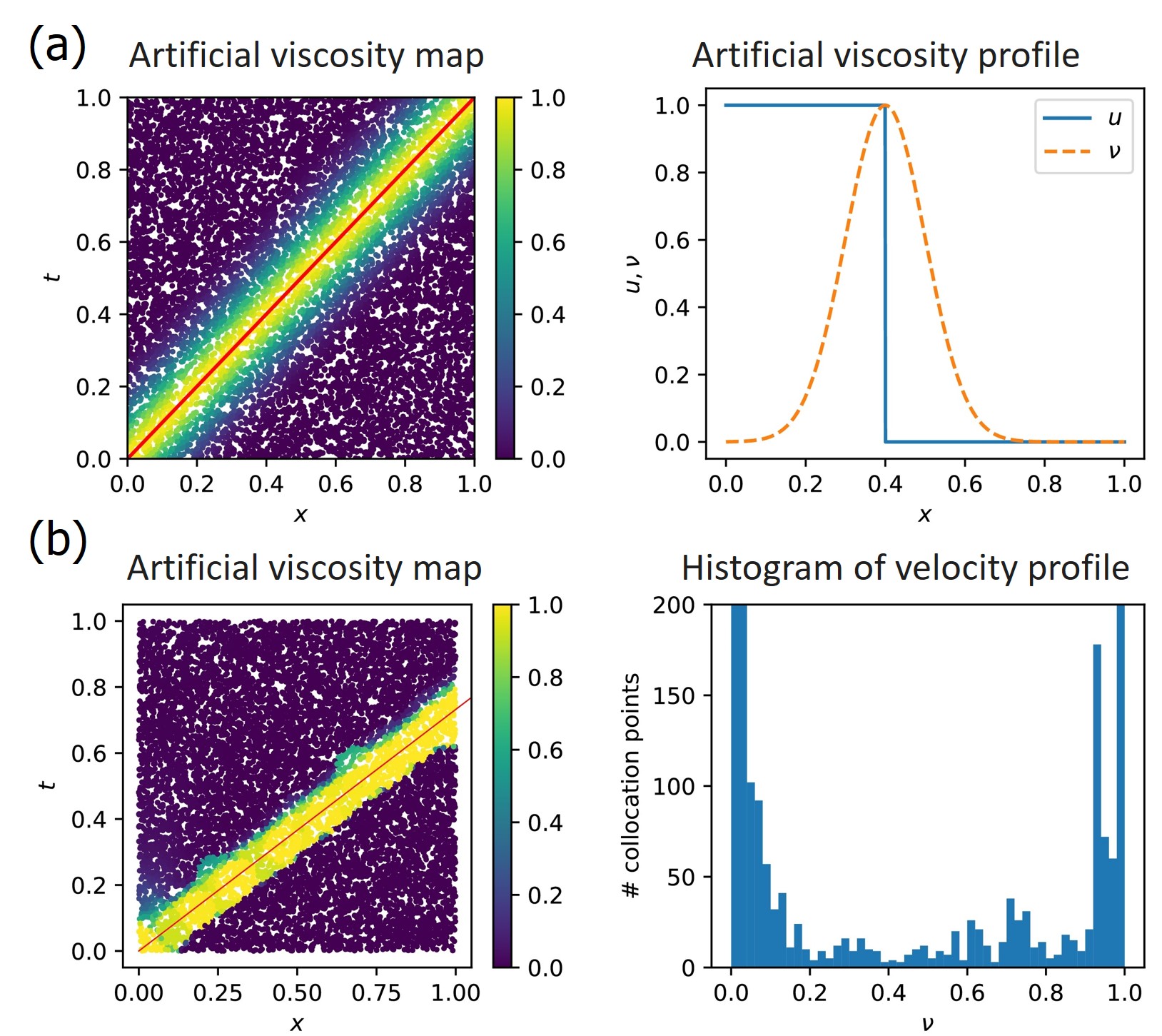}
    \caption{A visualization of two different techniques for defining adaptive artificial viscosity maps, based on \citet{Coutinho2023Physics-informedViscosity}: (a) parametric artificial viscosity map and (b) residual-based artificial viscosity map.}
    \label{fig:artificialviscositymap}
\end{figure}

These modifications mimic the effect of a capillary pressure term in the system, which adds stabilization by smoothing out the gradients in water saturation and pressure. The artificial viscosity method was then expanded in the later studies:

\begin{itemize}
    \item \textbf{Adaptive Artificial Viscosity.} Reaching reliable solutions for different system parameters was determined not to be possible using a unique and global value of \( \epsilon \). So,  \citet{Coutinho2023Physics-informedViscosity} proposed utilizing an adaptive value for \( \epsilon \) in which the value is learned adaptively during training of PINNs, based on the complexities in the solution of the system. In this technique, a trainable value for \( \epsilon \) is defined and a loss term based on that is added to the optimizer to find the smallest possible value for \( \epsilon \). 

    \item     \textbf{Parametric Localized Artificial Viscosity.} Applying a unique non-local value of \( \epsilon \) raised concerns about the accuracy of the final solution \cite{Huang2022AnScheme,PopovModifiedSystems,Albright2020LocallyHydrodynamics}. To address these concerns, \citet{Coutinho2023Physics-informedViscosity} explored the application of localized \( \epsilon \) values and showed improvements in the final solution of the BL equation. The main goal of this modification is to apply artificial viscosity locally, rather than across the entire computational domain. In this modification, the artificial viscosity is only applied in regions where discontinuities or shocks occur, leaving smooth areas unaffected. The modification involves constructing a parameter map that is informed by the prior information derived from the problem's solution, for instance spatiotemporal location of shock in a problem of multiphase flow. The equation \ref{eq:artificialviscosity2p} is subsequently updated to:

    \begin{equation}
    \frac{\partial s_w}{\partial t} + \frac{\partial f(s_w)}{\partial x} = \epsilon_{max} \epsilon (x,t) \frac{\partial^2 s_w}{\partial x^2},
    \end{equation}
    where $\epsilon_{max}>0$ is the maximum allowed amount of artificial viscosity, and $\epsilon (x,t)$ is the spatial-temporal artificial viscosity map that has its values bounded between 0 and 1. 
    This equation, requires an additional loss term to tune the $\epsilon_{max}$ value, that changes the total loss function to:
    
    \begin{equation}
    \mathcal{L}_t = \mathcal{L}_{\text{data}} + \mathcal{L}_{\text{physics}} + \mathcal{L}_{\text{initial}} + \mathcal{L}_{\text{boundary}} + \omega_{visc} \mathcal{L}_{\text{visc}},
    \end{equation}
        
    where ${\mathcal{L}}_{visc}(\nu_{\max}) = \epsilon_{\max}^2$. In the Buckley-Leverett problem, the exact position of the shock front can be determined using Eq. \ref{eq:shocklocationBL}, while the artificial viscosity distribution can be modeled using a Gaussian probability density function centered at the position of shock front, as illustrated in Fig. \ref{fig:artificialviscositymap}a. However, although promising, this method is limited to simple problems with predictable shock behavior.

    \item  \textbf{Residual-based Artificial Viscosity Map.} In this strategy, as proposed by \citet{Coutinho2023Physics-informedViscosity}, instead of relying on prior information about location and structure of shock, it is suggested to use a local map corresponding to the residual values of the PDE. This approach has been shown to be effective in tracking the location of shocks in more complex structures, as shown by \citet{Stiernstrom2021ALaws} for numerical methods. So, the updated PDE is written as,
    
    \begin{equation}
    \frac{\partial s_w}{\partial t} + \frac{\partial f(s_w)}{\partial x} = \epsilon_{\text{max}} \epsilon(x, t, \theta) \frac{\partial^2 s_w}{\partial x^2},
    \label{eq:adaptiveviscresidualmap}
    \end{equation}
    where, $\epsilon_{max}$ is a learnable parameter, representing the maximum limit of artificial viscosity, and $\epsilon(x, t, \theta)$ is regularizing factor, that depends on the residual values of the PDEs \cite{Stiernstrom2021ALaws}. The value is bounded with a maximum value of the residuals, and also a normalized value of the inviscid PDE residual, as detailed in \cite{Coutinho2023Physics-informedViscosity}. An example of applying this approach to the Buckley-Leverett problem is shown in Fig. \ref{fig:artificialviscositymap}b, where the residual locations are accurately tracked. This strategy is advantageous compared to previous methods, as it provides a largely unsupervised approach for accurately tracking and smoothing shock front effects, specifically for more complex shock behaviors.

\end{itemize}

\subsubsection{Total Variation Penalization}

The total variation (TV) of a continuous function measures the extent of oscillations over a spatial interval \([a,b]\). It is defined as the integral of the absolute spatial gradient of the 1D function \(u(x,t)\) \cite{Harten1983HighLaws}:
\begin{equation}
TV_a^b(t) = \int_a^b \left| \frac{\partial u(x,t)}{\partial x} \right| \, dx.
\label{eq:total_variation}
\end{equation}

In computational fluid dynamics (CFD), maintaining consistency in TV trends is crucial. For example, in the infinite-acting counter-current capillary diffusion problem, TV is expected to remain constant over time \cite{Abbasi2023SimulationNetworks}. However, when addressing shock wave problems, TV is expected to be constant or decrease over time (in 1D porous media flow, the total variation remains constant until the front reaches the outlet boundary, after which it decreases). To formalize this behavior, the concept of total variation diminishing (TVD) is introduced, which ensures:
\begin{equation}
TV(t_2) \leq TV(t_1), \quad \text{for } t_2 > t_1,
\label{eq:total_variation_dem}
\end{equation}
where the total variation at a later time step \(TV(u^{n+1})\) is less than or equal to the total variation at the current time step \(TV(u^n)\).
In the context of shock discontinuities, \citet{Patel2022ThermodynamicallySystems} proposed incorporating TVD as an additional training constraint to obtain thermodynamically valid solutions. For a one-dimensional system, this concept can be implemented as an auxiliary loss term:
\begin{equation}
\mathcal{L}_{TVD} = \sum_{n,p} \max\left(0, TV_h(u^p_{n+1}) - TV_h(u^p_n)\right)^2,
\label{eq:total_variation_loss}
\end{equation}
where \(TV_h\) represents the discrete approximation of TV. This loss term penalizes nonphysical oscillations in phase saturation during shock front propagation, promoting stability and accuracy in the solution. However, this approach alone may not fully resolve the errors at shock fronts and should be combined with other methods for improved effectiveness.

\subsubsection{Entropy Condition}

Entropy, a measure of disorder or randomness in a system, is a key concept in the analysis of shock waves. \citet{Lax1973HyperbolicWaves} introduced entropy conditions as necessary constraints for obtaining physically admissible solutions to quasilinear hyperbolic PDEs derived from conservation laws. In fluid flow systems, the entropy condition ensures compliance with the second law of thermodynamics by requiring that entropy increases as particles cross a shock front, reflecting the irreversibility of the process \cite{Liu2021ShockWaves}. The entropy conditions can be summarized as follows \cite{Fraces2021PhysicsMedia}:

\begin{itemize}
    \item \textbf{Entropy Condition:} On the downstream side of the shock (ahead of the front), front velocities should be less than or equal to both the shock velocities and the velocities on the upstream side of the shock \cite{Mishra2019NumericalEquations}.
    \item \textbf{Velocity Constraint:} Raised from the first condition, in a two-phase displacement process, the wave velocities must decrease steadily as upstream saturation (behind the front) increases.        
\end{itemize}

These conditions have been integrated into PINNs in several studies, such as \cite{Diab2022Data-FreeProblem, Fuks2020LIMITATIONSMEDIA, Zhang2024Physics-InformedSolubility, HuangONPREPRINT}, to constrain phase velocities by modifying the fractional flow profile at the shock front. For example, assuming the water saturation at the shock front is determined using the Welge graphical method, the fractional flow profile along the x-axis can be modified to restrict displacement to the convex hull of the fractional flow curve (for $s_w > s_{wf}$ in Fig. \ref{fig:welge}):

\begin{equation}
\tilde{f}_w(s_w) =
\begin{cases}
\frac{s_{w} - s_{wc}}{f(s_{ws})}  & \text{if } s_{wc} \leq s \leq s_{ws} \\
f_w(s_w) & \text{if } s_w > s_{ws}
\end{cases}
\end{equation}
where, $s_{ws}$ represents the water saturation at the shock front and is determined from the fractional flow curve using the Welge rule \cite{Welge1952ADrive}. Substituting this modified fractional flow into the Buckley-Leverett (BL) equation gives:

\begin{equation}
\ensuremath{  \frac{\partial {s_{w}}}{\partial t} + \frac{u}{\phi}\frac{\partial {\tilde{f}_w}}{\partial s_w}\frac{\partial {s_{w}}}{\partial x} = 0 ,\ \ 
\left(i=w,nw\right) }
\label{eq:BLeqent}
\end{equation}

Building on this, \citet{Diab2022Data-FreeProblem} proposed an alternative modification to the fractional flow:

\begin{equation}
\tilde{f}_w(s_w) =
\begin{cases}
    \sigma s_w, & \text{if } s_{wc} \leq s_w < s_{wf}, \\
    f_w(s_w), & \text{if } s_w \geq s_{wf}.
\end{cases}
\end{equation}
where, $\sigma$ is calculated as:
\begin{equation}
\sigma = f'(s_{wf}) = \frac{f(s_{wf})-f(s_{wi})}{s_{wf} - s_{wi}}  
\end{equation}

This approach has also been extended to problems involving two shock fronts \cite{Zhang2024Physics-InformedSolubility}.
\citet{Patel2022ThermodynamicallySystems} have suggested to use the entropy conditions to restrict the PINNs solutions across shock waves by defining a set of entropy flux pair, $ (\eta(u), q(u))$, in which the below condition is needed to be satisfied (apart from other constraints):
\begin{equation}
\partial_t \eta + \nabla \cdot q \leq 0
\end{equation}
This condition ensures thermodynamic consistency across shock fronts. Similarly, \citet{Jagtap2022Physics-informedFlows} applied entropy conditions, in combination with other techniques, to achieve viscosity solutions for steady-state compressible Euler equations.

\subsubsection{Weak Physics-Informed Neural Networks (wPINNs)}

The weak formulation of a partial differential equation (PDE) is a powerful mathematical tool that reformulates the original equation into an integral form. This approach is particularly beneficial for addressing problems with non-smooth solutions or irregular domains. Instead of enforcing the PDE to hold at every point in the domain, the weak formulation ensures that the equation is satisfied in an averaged sense.

To derive the weak form, the PDE is multiplied by a smooth test function, $\psi(x)$, and integrated over the domain $\Omega$. For the conservation law expressed as eq. \ref{eq:pdedef}, the weak form is written as:

\begin{equation}
\int_\Omega \left( \frac{\partial u}{\partial t} + \nabla \cdot f(u) \right) \psi(x) \, d\Omega = 0, \quad \forall \psi(x) \in \Omega,
\label{eq:wpinns_weakform}
\end{equation}

Using integration by parts, the divergence term is transferred to the test function, reducing the differentiability requirements for the solution $u(x,t)$:

\begin{equation}
\int_\Omega u \frac{\partial \psi}{\partial t} \, d\Omega - \int_\Omega f(u) \cdot \nabla \psi \, d\Omega = 0, \quad \forall \psi(x),
\label{eq:wpinns_diff}
\end{equation}

This transformation shifts the smoothness requirement from the solution $u(x,t)$ to the test function $\psi(x)$, making it possible to handle discontinuities in the solution. This property makes the weak formulation well-suited for problems involving shocks or abrupt changes in solution behavior \cite{Ambati2007SpacetimeEquations}.

In the context of PINNs, the weak formulation has been integrated into the framework to form Weak PINNs (wPINNs) \cite{DeRyck2024WPINNs:Laws, Zhang2024Weak-formulatedMaterials, Chaumet2024ImprovingSystems}. Unlike traditional PINNs, which enforce the strong form of the PDE, wPINNs leverage the weak form to enforce constraints. This is particularly useful for problems with discontinuities, such as shock waves in conservation laws.

For example, wPINNs have been successfully applied to solve non-linear conservation laws like the Burgers' equation \cite{DeRyck2024WPINNs:Laws}. Also, \citet{Chaumet2024ImprovingSystems} extended the wPINNs to other discontinuity problems, such as rarefaction waves and compressible Euler equations. The results report the capability of wPINNs in capturing entropy solutions, which are critical for ensuring the physical validity of the results. In these cases, the weak formulation aids in addressing the challenges posed by discontinuities, allowing for more robust approximations. However, this comes at the cost of increased computational complexity due to the need for numerical integration and the additional overhead of weak constraint implementation \cite{DeRyck2024WPINNs:Laws}. Additionally, the applications remain limited to simple 1D problems, highlighting the need for further investigation, e.g., for 3D problems.

\subsubsection{Change of Variables}
\begin{figure}
    \centering
    \includegraphics[width=0.9\linewidth]{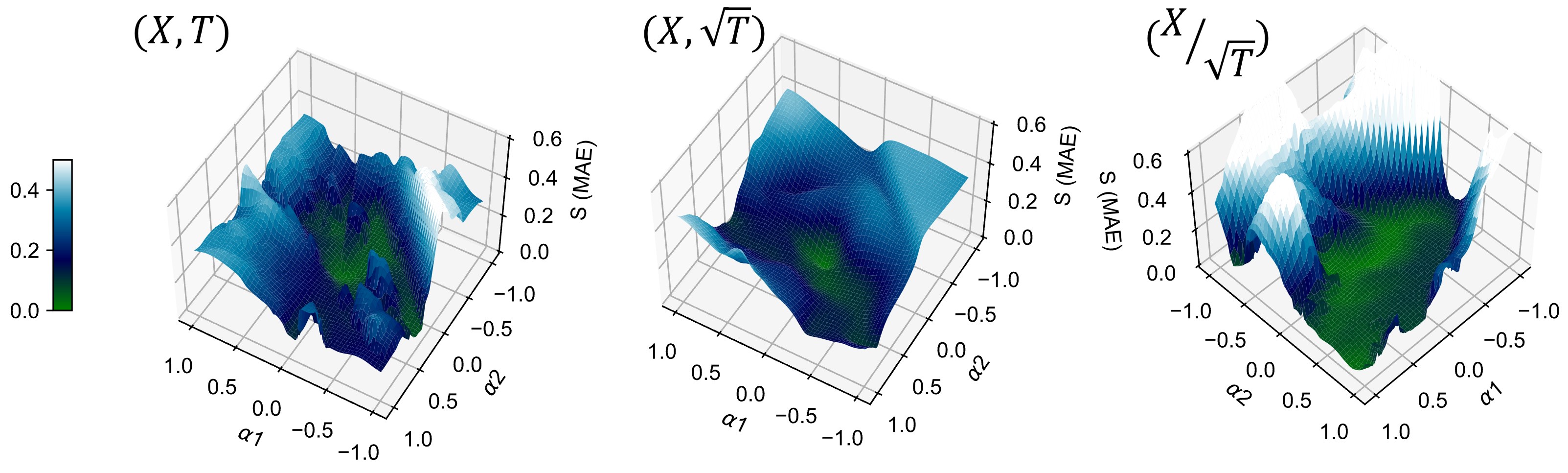}
    \caption{A comparison of the impact of change-of-variables technique in convexifying the loss landscape \cite{Abbasi2023SimulationNetworks}.}
    \label{fig:losslandscapechangeofv}
\end{figure}

The change of variables is a mathematical technique used to transform a problem’s variables, to reduce the complexity of PDEs, and enhancing solution accuracy \cite{Wu2024VariableProblems}. \citet{Abbasi2023SimulationNetworks} demonstrated that transforming variables into more physically informed representations can improve the accuracy of PINNs and convexify the loss landscape, as illustrated in Fig. \ref{fig:losslandscapechangeofv}. This approach is particularly effective in enhancing PINNs’ performance when dealing with discontinuities, such as rarefaction waves. In the benchmark section, we also show the possibility of using this approach for the problem of two-phase flow in porous media.

\subsection{Loss and Training Modification Methods}

Loss and training modifications enhance PINNs' capability to address discontinuities by introducing regularization to the training loss function. Techniques such as PINNs-WE and GA-PINNs adjust residual weighting in critical regions to improve shock resolution. Advanced sampling strategies further refine focus on regions with complex solution dynamics. The following sections provide a detailed review of these methods.

\subsubsection{Weighted-equations PINNs (PINNs-WE)} 
In this method that was first proposed by \citet{Liu2024DiscontinuityNetworks}, the focus was local manipulation of the residual loss, so that the high-gradient regions get lower weights. The weighting of the residual points is done by multiplying a factor that is formulated in a way that, 
\begin{equation}
\lambda_{WE} = \frac{1}{1 + \alpha \left( |\nabla  \mathbf{u}| - \nabla  \mathbf{u} \right) }
\label{eq:weno_weight}
\end{equation}
where $\alpha$ is the adjusting factor. The above equation is originated from the intuitive sense that the shock appears when $\nabla \mathbf{u} \rightarrow \pm \infty $. By applying this method, the PINNs-WE approach effectively reduces the distortion of the solution caused by sharp gradients or discontinuities. In another study, \citet{Neelan2024Physics-informedStudy} applied a similar approach (in combination with artificial viscosity) to address discontinuities in problems such as the Sod shock tube. They demonstrated the superiority of the PINNs-based method over traditional CFD models in terms of solution accuracy. However, this improvement came at the expense of significantly higher computational costs.
\subsubsection{Gradient-Annihilated PINNs (GA-PINNs)} 

A similar loss modification method was recommended by \citet{Ferrer-Sanchez2024Gradient-annihilatedHydrodynamics}, called Gradient-Annihilated PINNs, in which the model learns to ignore the errors in the regions with high gradients. The authors approached solving the shock waves in Riemann problems (referred to the Euler equation with a piecewise initialization). In this method, an adaptive multiplier is defined in which modifies the local residuals of PDEs, i.e., $\mathcal{R}_{\text{int}}(x, t)$, 
\begin{equation}
\lambda_{GA}(x,t) := \sum_{k=1}^{D} \frac{1}{(1 + \alpha_k |\partial_x u|)^{\beta_k}}
\label{eq:gradient_anh}
\end{equation}
where $\alpha_k$ and $\beta_k$ are the hyper-parameters of the multiplier $\lambda_{GA}(x,t)$, determining the impact level of the multiplier. Then, the PDE residual in eq. \ref{eq:residuals} can be rewritten as:
\begin{equation}
\mathcal{R}_{\text{GA-D}}(x, t) := \lambda_{GA}(x,t) \mathcal{R}_{\text{D}}(x, t),
\label{eq:residualsGE}
\end{equation}

The authors also highlighted the significant influence of activation functions on the overall performance of the calculations, noting that certain functions may enhance or impede the training process. While the proposed methodology shows promise in addressing the challenges posed by discontinuities in PINNs, the authors emphasized the need for further research on automating hyperparameter selection to optimize performance. This includes refining methods to dynamically adjust key parameters during training to improve robustness and efficiency in handling complex solutions.

\subsubsection{Advanced Collocation Point Sampling} 
\begin{figure}
    \centering
    \includegraphics[width=0.7\linewidth]{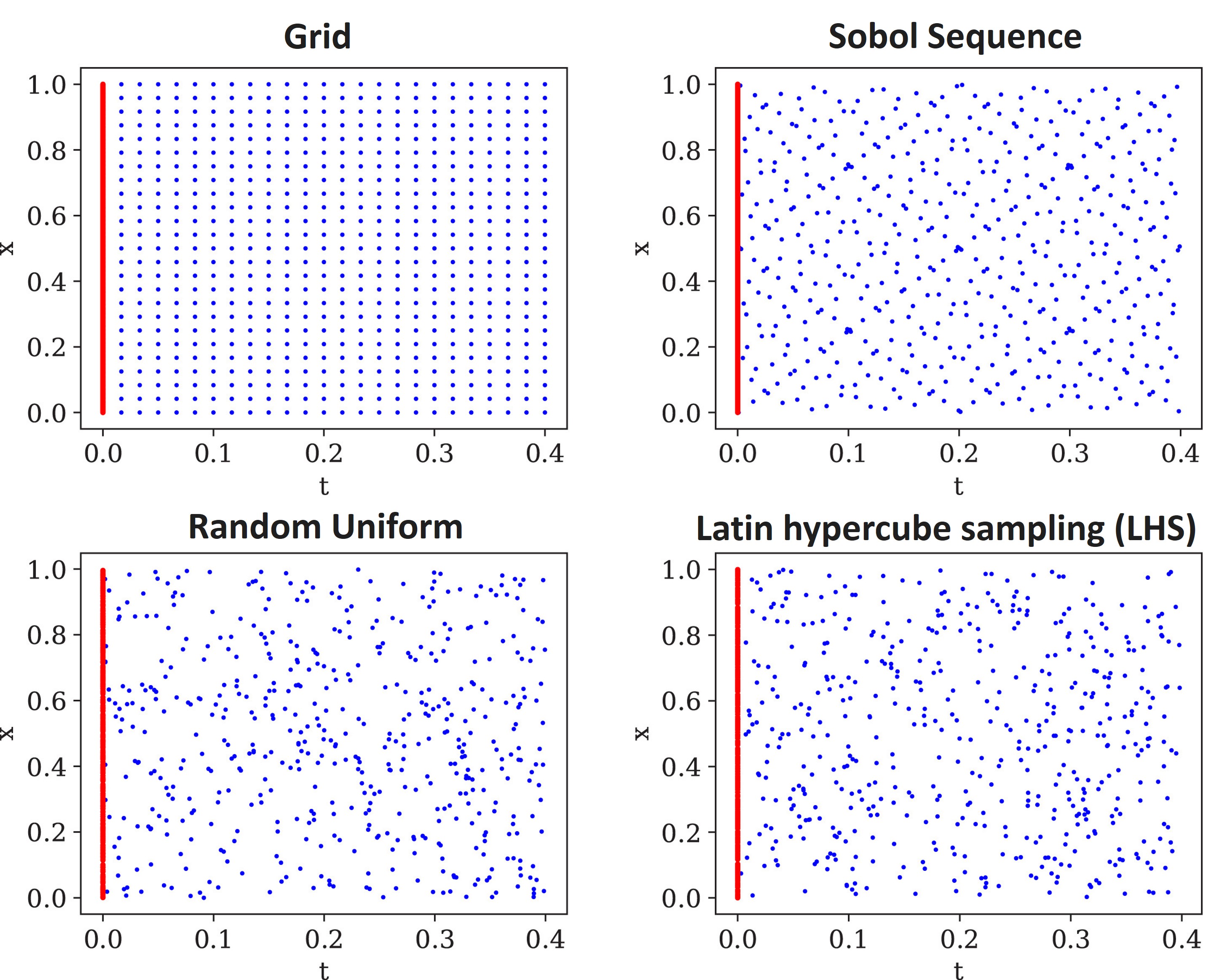}
    \caption{A comparison of various collocation point sampling strategies, illustrating the distribution of 500 generated points in the solution domain. The initial domain points, corresponding to $t=0$, are shown in red, while the internal collocation points, where the residuals of the PDE system are calculated, are shown in blue \cite{Ferrer-Sanchez2024Gradient-annihilatedHydrodynamics}.}
    \label{fig:collocationpointsampling}
\end{figure}

Several studies have demonstrated the positive impact of employing innovative collocation point sampling techniques on the generalization performance of PINNs \cite{Wu2023ANetworks, Zeng2022AdaptiveEquations, Chen2024AdaptiveMethods}. Various sampling strategies have been proposed in this context, including uniform sampling, random sampling, moving sampling \cite{Yang2024MovingPDE}, Sobol sequences, and residual-based adaptive refinement with distribution (RAR-D) \cite{Wu2023ANetworks}, which are among the most validated approaches. A comparison of various collocation point sampling strategies is illustrated in Fig. \ref{fig:collocationpointsampling}. In specific, residual-based adaptive sampling strategies focus the sampled points on regions with higher optimization errors, which can be particularly useful for problems with localized complexities like shocks \cite{Hanna2022Residual-basedNetworks, Mao2023Physics-informedSolutions}. While modifying sampling strategies alone may not completely overcome the challenges of accurately capturing shock fronts, these methods—especially when combined with other techniques, as discussed in  \citet{Ferrer-Sanchez2024Gradient-annihilatedHydrodynamics}—can enhance the overall performance of PINNs in complex scenarios, such as those involving shock waves or rarefaction waves.

\subsection{Architecture modification methods}
Architecture-based methods focus on enhancing the flexibility of the deep learning models used as the backbone for PINNs solvers. These techniques aim to develop strategies that improve the networks' ability to capture strong nonlinearities in the system of equations. While numerous techniques can be applied to equations with rapid value changes, this section focuses specifically on methods designed to enhance PINNs performance in handling shock fronts.

\subsubsection{Transformer Based Architecture}
Inspired by the attention mechanisms introduced by \citet{Vaswani2017AttentionNeed}, \citet{Wang2020UnderstandingNetworks} proposed an architecture in which the PINNs performance improved across different benchmark problems. In this architecture, the input variables are passed through two transformer networks, $U$ and $V$, which project the inlet variables into high-dimensional feature space. Then, the outlets of each latent layer in the networks are updated via a pointwise multiplication, written as:

\begin{equation}
H^{(k)} = (1 - H^{(k)}) \odot U + H^{(k)} \odot V, \quad k = 1, \dots, L
\label{eq:transform}
\end{equation}
where  $H^k$ denotes the $(n \times h)$  latent matrix of MLP network at layer $k$ , and  $\odot$ denotes elementwise multiplication (see Fig. \ref{fig:archcomp}). The authors found the applied modification improves the performance of PINNs because of two characteristics: 1) Specifically addresses the multiplicative interactions among various input dimensions, and 2) Strengthens the hidden states using residual connections. The authors also reported the improvements in the performance of PINNs in terms of convergence rate, accuracy, and stability. Then, the technique was applied for calculating the problem of viscous flow in porous media with shock fronts by \citet{Diab2022Data-FreeProblem}. They showed the transformer-based deep learning technique, the same as in eq. \ref{eq:transform}, can overcome the problem of capturing discontinuities in Buckley-Leverett type problem. The approach in general showed improvements in the performance of PINNs and neural operators in \cite{Anagnostopoulos2024Residual-basedNetworks}. \citet{Li2024AnLearning} integrated the attention mechanism into the Long Short-Term Memory (LSTM) modeling framework to address discontinuities in shallow water equations using PINNs.
A modified version of using attention mechanisms also was used in the work of \citet{Rodriguez-Torrado2022Physics-informedProblem}. In a similar study, \citet{Dong2023SolvingPINNs} utilized skip connection transformer to mitigate the gradient vanishing problem in PINNs when solving an ideal magnetohydrodynamics problem (Dai-Woodward shock tube problem), which is characterized by various types of discontinuities.

\subsubsection{Extended PINNs (XPINNs)}
Extended PINNs (XPINNs) is a technique that proposes generalized spacetime domain decomposition approach for solving PDEs via PINNs on arbitrary complex/large geometry domains \cite{Jagtap2020ExtendedEquations}. \citet{Jagtap2022Physics-informedFlows} proposed the application of XPINNs for the solution of the inverse problems related to compressible supersonic flow, in parallel to using other relevant techniques such as adaptive activation functions \cite{jagtap2020adaptiveAF}, dynamic weights \cite{Wang2020UnderstandingNetworks}, and entropy conditions \cite{Godlewski1996NumericalLaws}. Supersonic flows often exhibit various types of discontinuities in the solution, including oblique shock, expansion waves, and bow shock waves. These discontinuities pose significant challenges for traditional numerical methods, as well as the vanilla PINNs. While the paper shows promising results for spatially stationary shocks inverse problems, the application of XPINNs to dynamically moving shocks problems would require further research and development.

\subsubsection{Fourier Features}
Fourier features have been shown to improve the flexibility of neural networks to model high-frequency functions by mapping inputs to a higher-dimensional space via sinusoidal transformations \cite{Tancik2020FourierDomains}. Several approaches to incorporating Fourier transformations into PINNs have been proposed in the literature, including Fourier feature embeddings \cite{Wang2020OnNetworks} and Fourier encoders/decoders \cite{Abbasi2023SimulationNetworks}. This mapping helps mitigate the smoothness (spectral) bias of standard neural networks, enhancing their capability to capture high-gradient regions in PDEs. However, it may still present challenges when applied to shock waves.

\subsubsection{Relaxation Neural Networks}
According to the approach proposed by \citet{Zhou2024CapturingNetworks}, the use of composite relaxation neural networks can increase the accuracy of PINNs in capturing shock fronts by introducing local relaxation approximators. In this approach, separate networks approximate the accumulation ($s_w$), and flux ($f_w(s_w)$) in the conservation laws. For instance, for eq. 
\ref{eq:systemof1d2pflow}, we may define two different neural networks, i.e. $\mathcal{N}^{\theta}(s_w^{\theta}(\mathbf{x}, t))$, and $\mathcal{N}^{\theta}(f_w^{\theta}(\mathbf{x}, t))$, in such a way that $ f_w(s_w^{\theta})=f_w^{\theta}(s_w)$. Then, we need to solve the following system of equations,

\begin{equation}
\begin{cases}
\frac{\partial {s_{w}^{\theta}}}{\partial t} + \frac{v}{\phi}\frac{\partial {f_{w}^{\theta}}}{\partial x} = 0 \quad \\
f_w^{\theta} - f_w(s_w^{\theta}) = 0
\end{cases}
\end{equation}

This implies that a residual term, serving as a dissipative correction, must be defined during the training process to account for errors in the local approximations of fractional water mobility, 
\begin{align}
\mathcal{R}_{\text{relax}}(x, t) = f_w^{\theta} - f_w(s_w^{\theta}),
\label{eq:relaxloss}
\end{align}

The proposed technique was demonstrated for different problems such as the Euler equations, and Shallow water equations, and Burgers’ equation. \citet{Zhou2024CapturingNetworks} also reported that the method can be used in combination with other shock management techniques.

\subsubsection{Lagrangian PINNs (LPINNs)} \citet{Mojgani2023KolmogorovPDEs} introduced Lagrangian PINNs (LPINNs) to address failure modes in vanilla PINNs in the problems involving transport phenomena. LPINNs reformulate the problem in the Lagrangian framework, i.e., $(\xi,t)$ - instead of the Eulerian framework, i.e., $(x,t)$, in vanilla PINNs - to align the causality of the problem more carefully. A schematic of the architecture of LPINNs compared to vanilla-PINNs is illustrated in Fig. \ref{fig:archcomp}. The authors found that the moving discontinuities in the Eulerian domain generate causality issues, while LPINNs solve this problem by aligning the neural network's architecture with the natural causality of the system. In this approach, two parallel networks are designed in which one ($N_u$) is responsible for the state variables, and the other one ($N_x$) is responsible for the spatial coordinates of the system. The proposed method seems to improve the performance of PINNs in solving a series of challenging examples such as the viscous Burgers equation and convection-diffusion.

\subsubsection{Cusp Capturing PINNs} 
\begin{figure}[htp]
    \centering
    % \includesvg[width=14cm]{figs/figarch.png}
    \includegraphics[width=15cm]{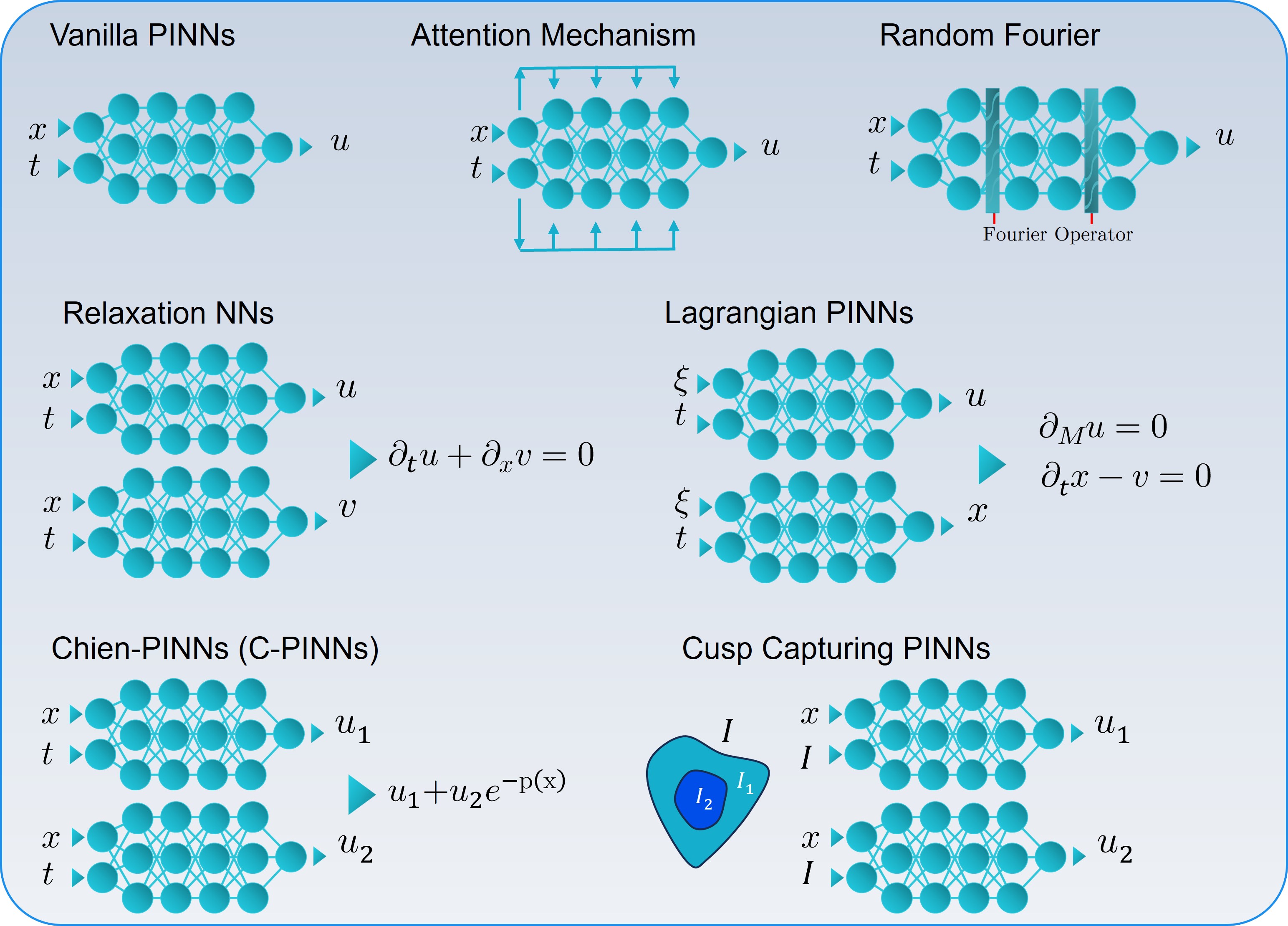}
    \caption{A schematic comparison of the different network architectures proposed for improving the PINNs performance in capturing the shock fronts efficiently.}
    \label{fig:archcomp}
\end{figure}

\citet{Tseng2023AForces} proposed handling discontinuities in Stokes interface problems (static discontinuity) using PINNs with a specialized structure designed to capture both discontinuities and cusps. To address the discontinuous pressure and the derivative discontinuities of velocity at the fluid interface, the authors employed two separate networks: one for pressure (handling function discontinuity) and another for velocity (handling derivative discontinuity). Each sub-network utilizes an augmented feature, where a level-set ($I$) function encodes the position and/or properties of the domain (Fig. \ref{fig:archcomp}). This approach allows the PINNs to maintain the physical properties of the solution, effectively capturing the discontinuous and cusp-like behaviors at the interface, thus offering an alternative to traditional grid-based methods, such as the immersed interface method, while achieving comparable accuracy. Later, \citet{Li2025TwoSurfaces} expanded the methodology to the more complex problems of elliptic interface problems. This method works by explicitly defining discontinuities through solution decomposition ($u=u_{I_1}+u_{I_2}$), avoiding the network's struggle with high-gradient values across the discontinuity interface. Similarly, \citet{Abbasi2024History-MatchingPINNs} employed a domain decomposition approach to model flow in fractured porous media. By separately solving the PDEs in the low-frequency (rock) and high-frequency (fracture) domains, the method avoids spectral bias around discontinuities (fractures), enabling PINNs to better capture the solution in both domains.

\begin{figure}
    \centering
    \includegraphics[width=0.9\linewidth]{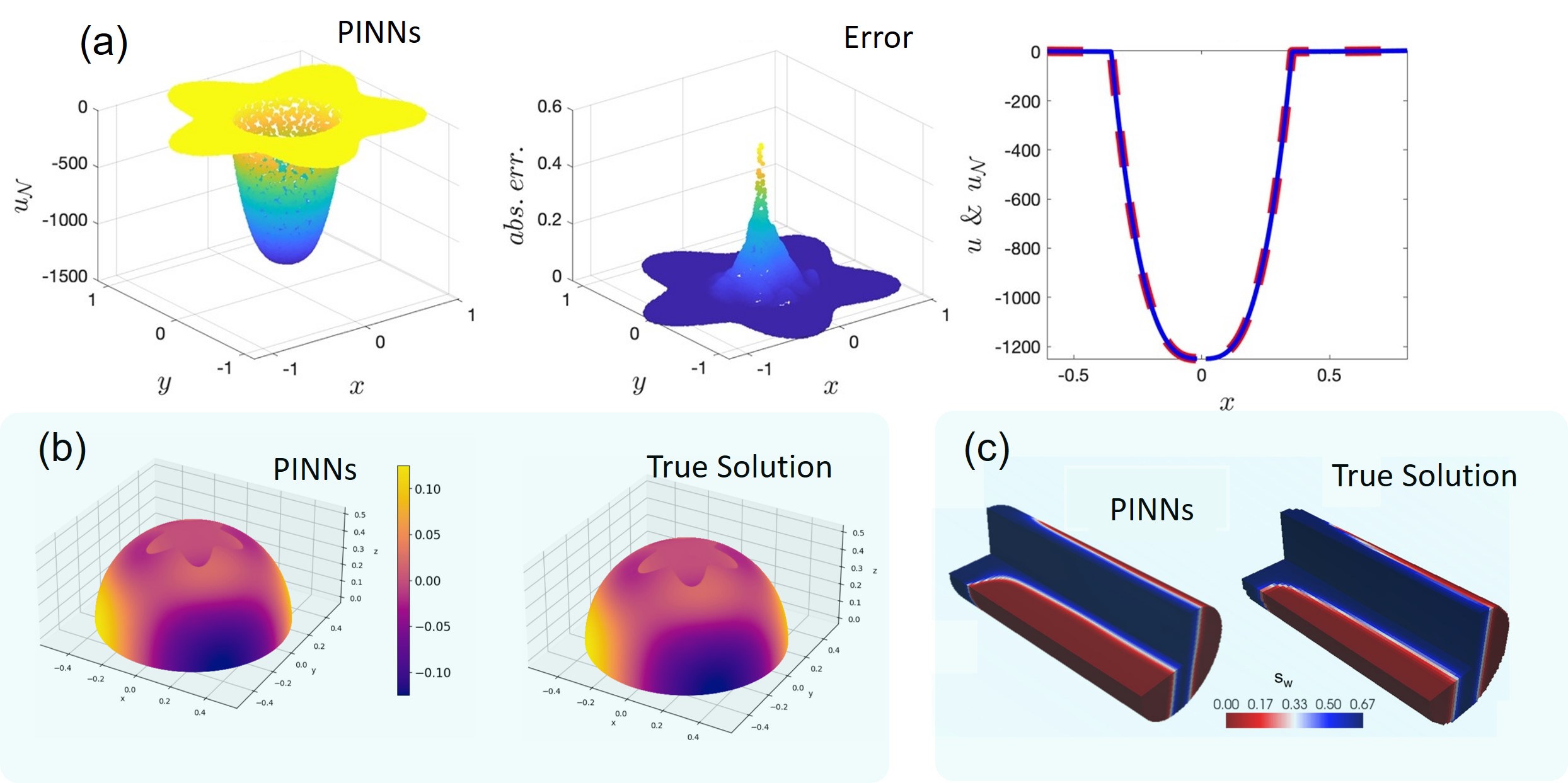}
    \caption{Examples of using PINNs for Cusp-like problems. (a) Cusp-capturing PINNs applied for solving an interface problem defined on an irregular domain (cusp discontinuity) \cite{Tseng2023AForces}, (b) Solving an elliptic interface problem on a half-sphere \cite{Li2025TwoSurfaces}, (c) Using PINNs for capturing extreme discontinuities in problem of flow in fractured porous media \cite{Abbasi2024History-MatchingPINNs}. }
    \label{fig:cusp}
\end{figure}

\subsubsection{Activation Functions} 
In several works it has been demonstrated that the optimal selection of activation functions can improve the overall performance of PINNs in reducing both approximation and generalization errors \cite{DeRyck2022ErrorPDEs, Jagtap2023HOWDIRECTIONS,Abbasi2024PhysicalPINNs}. For instance, \citet{jagtap2020adaptiveAF,jagtap2022deepKNN,jagtap2020locally} introduced adaptive activation functions, which incorporate a trainable parameter to dynamically adjust the nonlinearity of the activation function and overall representativity of the network during the training process. This approach allows the network to modify the degree of nonlinearity based on the complexity of the problem, enabling better adaptation to challenging features such as sharp gradients or discontinuities. Their results - for instance, in the case of Burgers equation- demonstrated that adaptive activation functions can enhance the training dynamics by achieving faster convergence and improving accuracy in complex problems. In various other studies, such as \cite{Liu2024DiscontinuityNetworks}, activation functions have been employed as an effective strategy to improve the network's ability to handle sharp discontinuities in the solution.

\subsubsection{More architectures} 
In recent years, several studies have introduced alternative architectures that directly or indirectly can enhance the performance of PINNs for stiff PDEs. For instance, domain decomposition-based methods, such as Finite-Basis PINNs (FBPINNs) introduced by \citet{Moseley2023FiniteEquations} and Conservative PINNs (CPINNs) introduced by \citet{Jagtap2020ConservativeProblems}, can potentially improve PINNs performance in capturing high-frequency components of solutions. Other architectures, like Binary PINNs \cite{Liu2024BinarySolutions}, Physics-Informed Generative Adversarial Networks (PIGANS) \cite{Ma2024Physics-InformedEquation} may also prove beneficial in similar contexts

\begin{figure}
    \centering
    \includegraphics[width=0.99\linewidth]{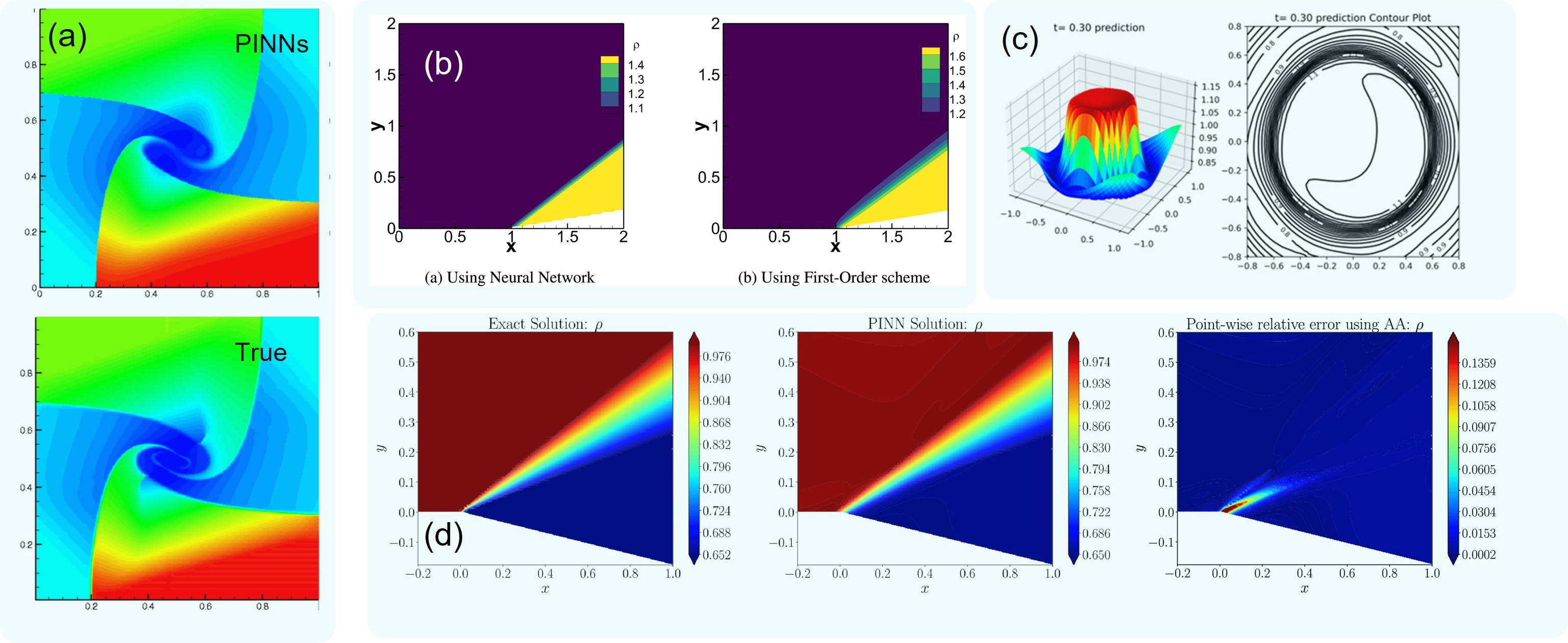}
    \caption{Examples of applying PINNs to the problems with discontinuities. (a) Solution of PINNs-WE model for a 2D Riemann problem \cite{Liu2024DiscontinuityNetworks}; (b) Using WENO-PINNs to solve non-linear Euler equation (gas dynamics), (c) Using an LSTM network architecture to capture the discontinuities in a two-dimensional dam-break problem using a Hyperbolic PDE \cite{Li2024AnLearning}, (d) PINNs applied to solve the expansion wave problem (rarefaction wave), using XPINNs domain decomposition approach \cite{Jagtap2022Physics-informedFlows}. }
    \label{fig:examplepinns}
\end{figure}

\section{Benchmarking}

In this section, we compare the performance of the introduced approaches across different levels of complexity in the underlying PDE. We begin with the Buckley-Leverett (BL) equation (constant flow velocity), as a special case of the two-phase flow equations. Then, we focus on the performance of PINNs for the coupled 1D two-phase flow equation system of equations (constant pressure drop). To evaluate different strategies, we selected a range of techniques from each of the three categories for a comprehensive evaluation. A summary of the properties of the system under study is shown in Table \ref{table:systemprops}.

\begin{table}[ht]

\caption{System properties in the benchmarking case study.}
\begin{center}
    \label{table:systemprops}
    \small 
    \begin{tabular}[t]{llll}
    \hline
    Property & Value & Property & Value  \\ \hline
    $\phi$ (-)  & 0.10 & \(k_{rw}^{\max}\)   &  1.0 \\ \hline       
    $k$ ($m^2$)  &  1e-16 & \(k_{rnw}^{\max}\) & 1.0 \\ \hline
   \(n_{w}\)  & 2.0 & \(n_{nw}\)    &   2.0 \\ \hline
    \(\mu_w\), \(\mu_{nw}\) (cP)  &  1.0, 1.0 & IFT $(N/m)$ & 0.00 \\ \hline
     \(s_{wc}\) & 0.0  &  \(s_{nwr}\) & 0.0 \\ \hline

    \end{tabular}
\end{center}
\end{table}
\textbf{}

\subsection{Two-Phase 1D Flooding in Porous Media at Constant Injection Rate: The Buckley-Leverett Problem.}
\begin{figure}[h!]
    \centering
    \includegraphics[width=13cm]{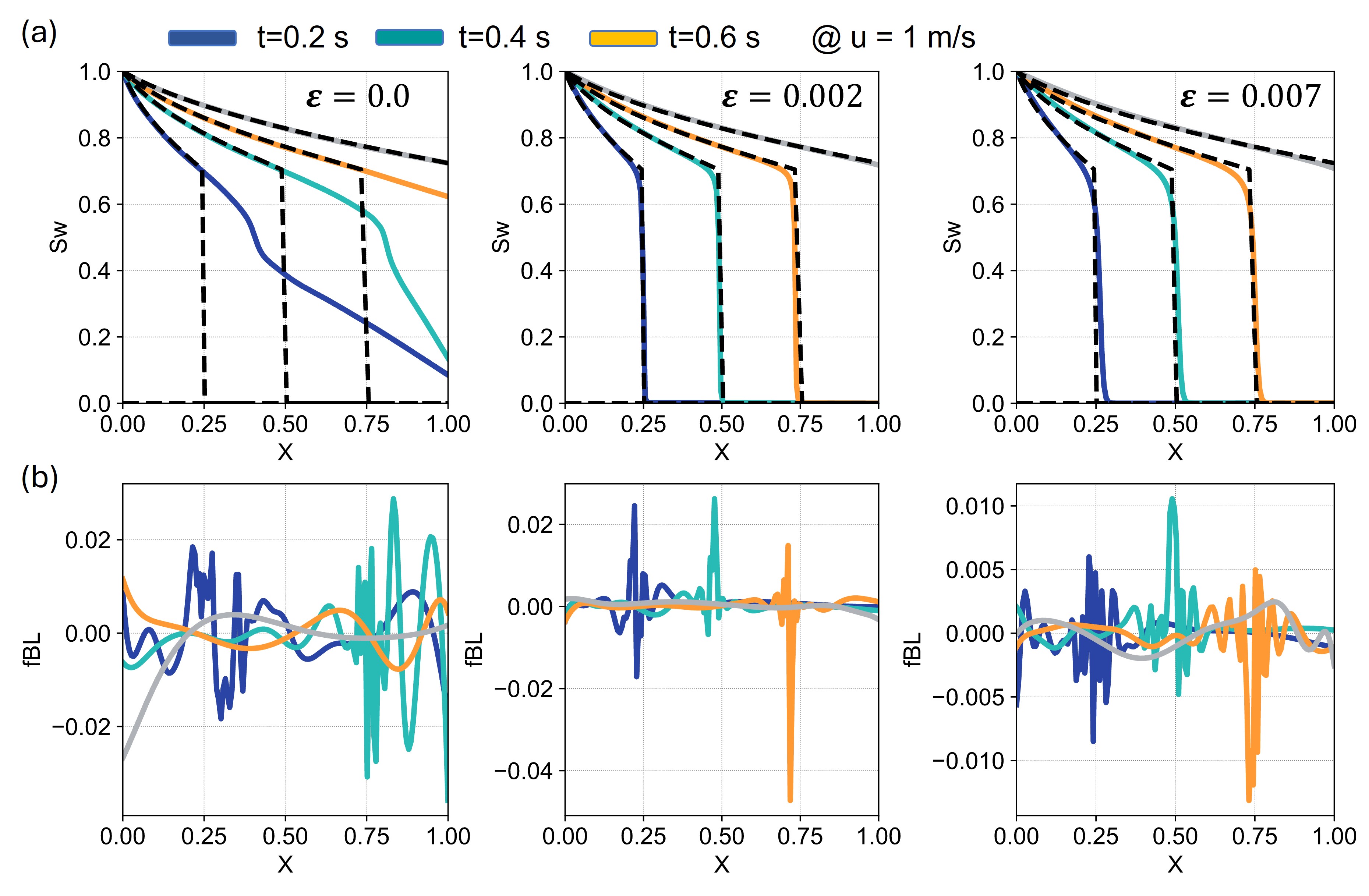}
    \caption{The performance of PINNs in solving the Buckley-Leverett equation of flow with different artificial viscosity coefficients. A standard Cartesian collocation point sampling strategy has been used, with 100 points in temporal and 200 points in spatial coordinates; (a) The front of water saturation at the different times, (b) the residual of PDE at the different times.}
    \label{fig:figbl}
\end{figure}

In this section, we evaluated various shock management strategies to address the challenge of capturing the two-phase front in the Buckley–Leverett (BL) equation, i.e., eq. \ref{eq:BLeq}. Table \ref{table:systemprops} presents the applied system properties. Here, the constant injection velocity of 1 m/s were assumed.  Although the BL equation is a simplified version of the eq. \ref{eq:systemof1d2pflow}, the issue of exploding gradients remains a significant challenge. Previous studies have demonstrated that vanilla-PINNs struggle to solve this problem, as illustrated in Fig. \ref{fig:figbl} a \cite{Fuks2020LIMITATIONSMEDIA, Fraces2021PhysicsMedia}. While the PINNs solution can accurately capture the saturation distribution behind the front, it fails to do so at the front itself. Consequently, the predicted trend of average saturation ($\bar{s_w}$) is also inaccurate, leading to significant errors in the mass-balance of the system. The residual trend along the x-coordinate exhibits significant oscillations, indicating the difficulties PINNs face in minimizing errors in the frontal region, as shown in Fig. \ref{fig:figbl} b. 

Table \ref{table:BL_compare} presents a comparison of several shock management methods for addressing the shock front problem in the BL equation, along with their respective computational times. In the vanilla-PINNs model, a fully-connected neural network with a $\tanh$ activation function was employed. The network architecture consisted of 4 hidden layers (depth) and 40 neurons per layer (width). For the collocation points, a Cartesian grid strategy was applied, utilizing 100 points in the temporal dimension and 200 points in the spatial dimension. The model was trained using the Adam optimizer with a learning rate of $1 \times 10^{-4}$. For the other cases, the vanilla-PINNs model changed accordingly. In all the cases, hyperparameter testing was performed to make sure the best possible solutions are obtained. 

\begin{figure}[h!]
    \centering
    \includegraphics[width=11cm]{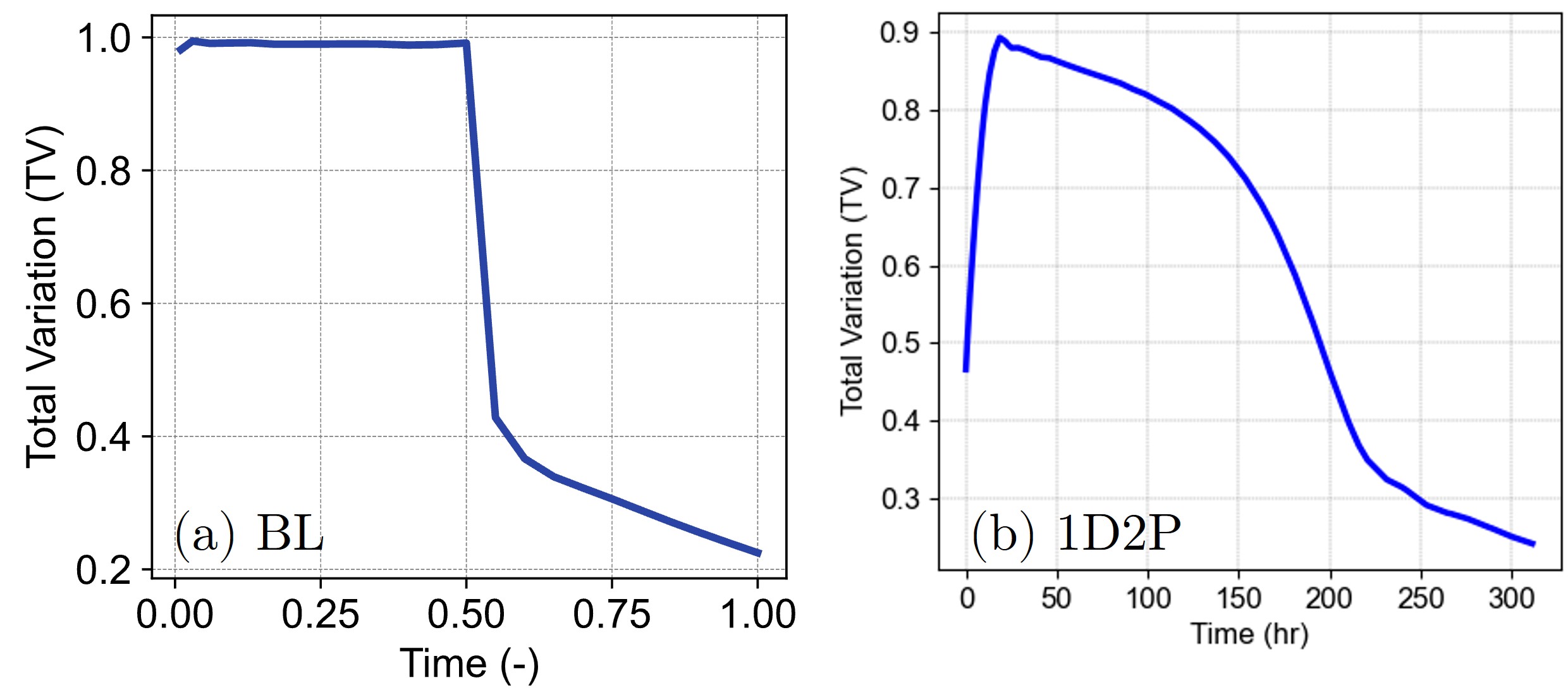}
    \caption{The trend of changes in TV versus time for Vanilla-PINNs model, in the solution of (a) BL-type PDE and (b) 1D two-phase flow PDE.}
    \label{fig:tvdtrack}
\end{figure}

\begin{figure}[h!]
    \centering
    \includegraphics[width=0.9\linewidth]{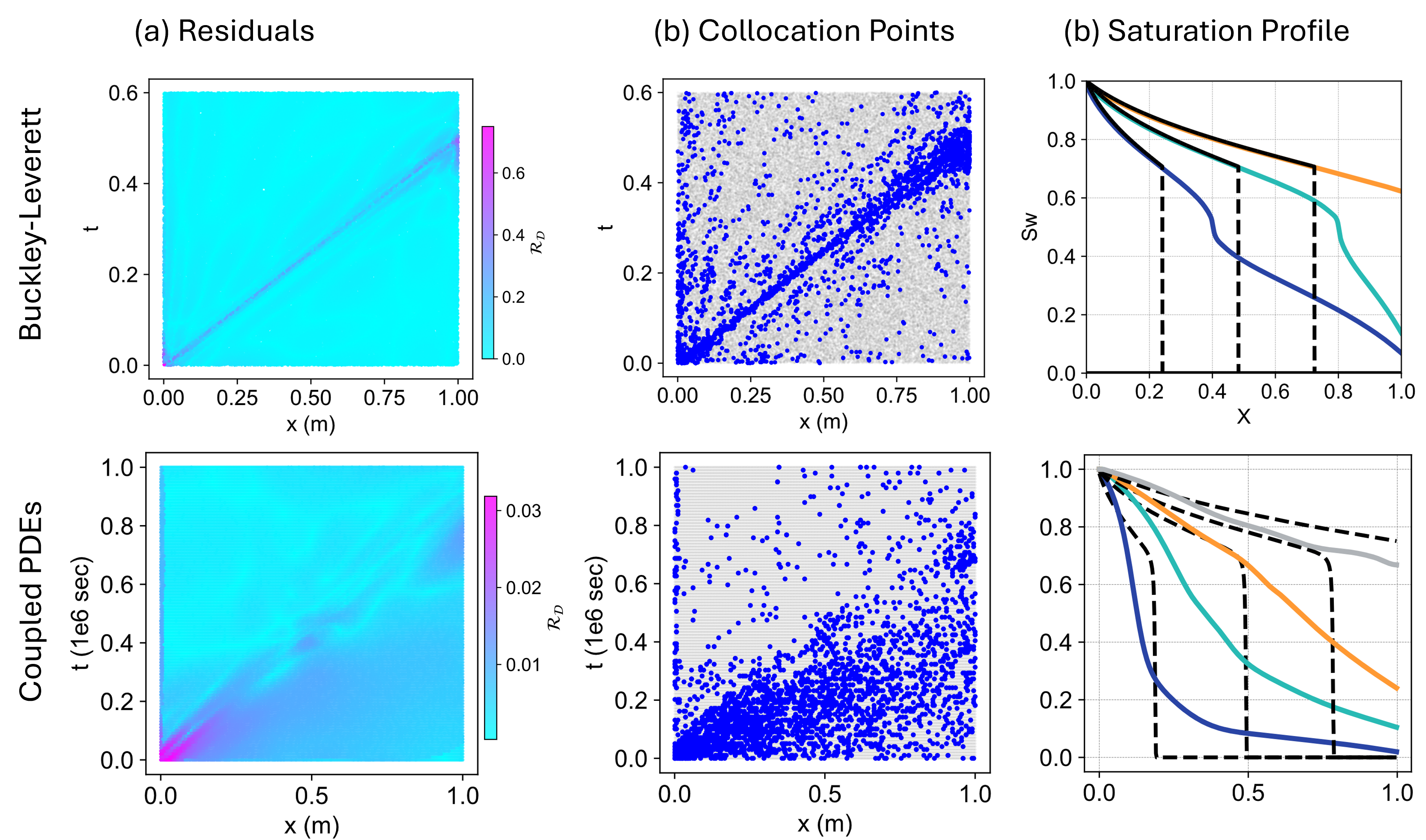}
    \caption{A visualization of the performance of residual based adaptive collocation point refinement \cite{Wu2023ANetworks} applied to both benchmark problems, a) the distribution of PDE residuals, b) the additional collocation points chosen based on the RAR-D strategy, c) the final solution of PINNs.}
    \label{fig:rac}
\end{figure}

It is clear that none of the architecture-based techniques could recover the solution with desired accuracy, with $s_w$ mean absolute error (MAE) of around 0.2. However, it looks physics-based methods could give reliable solutions, reducing the generalization error to a reliable value (with an $s_w$ MAE of around 0.02). Similarly, the loss regularization techniques could solve the shock issue, with acceptable generalization errors. The adaptive sampling of collocation points (RAR-D), aimed at increasing the density of points in regions with higher residuals, was also ineffective in resolving the issue, as demonstrated in Fig. \ref{fig:rac}. Additionally, the calculation of total variation (TV), as shown in Fig.~\ref{fig:tvdtrack}a, demonstrates that the vanilla-PINNs model adheres to the expected behavior of TV in the propagation of water saturation. Specifically, the TV remains either constant or decreases over time, as required, with the exception of negligible deviations during the initial stages of the flow. This indicates that incorporating TV constraints would not contribute to reducing computational errors in this context.
Although methods such as adaptive activation functions and Fourier transformations contributed to improving network flexibility, they did not resolve the issue of infinite gradients at the fronts.

The results confirm the relative advantage of physics-based and training-based methods for improving the performance of PINNs in modeling the problems with shock fronts. In the next section, this study will be extended to the full form of the PDE governing two-phase flow in porous media, which involves coupled conservation law equations for both wetting and non-wetting phases. These equations are more similar to the form commonly solved in real-world applications such as reservoir simulators.

\begin{table}[ht]
\caption{A comparison of the performance of different methods for forward simulation of Buckley-Leverett type PDE. The methods with the lowest errors in the predicted $s_w$ trend are highlighted.}
\begin{center}
    \label{table:BL_compare}
    \begin{tabular}[t]{llccc}
    \hline
    \textbf{Group} & \textbf{Method} & \makecell[c]{\textbf{Run-time} \\ (ms/ep.)} & \makecell[c]{\textbf{$L_t$ } \\ (MAE)} & \makecell[c]{\textbf{$s_w$ } \\ (MAE)} \\ \hline
   Vanilla PINNs & - & 15.8  & 5.6e+01  & 0.23  \\ \hline
    \multirow{2}{*}{Physics-Based} 
                    & \textbf{Artificial Viscosity} & \textbf{29.3}  & \textbf{7.1e+0}  & \textbf{0.02}  \\
                    & \textbf{Entropy} & \textbf{23.2}  & \textbf{1.2e+0} & \textbf{0.007}  \\                              \hline
    \multirow{1}{*}{Training} 
                    & \textbf{WENO} & \textbf{21.8}  & \textbf{4.1e+0}  & \textbf{0.028}  \\ 
                    & RAR-D & 18.1   &  1.5e+3  &  0.23  \\ \hline                            
    \multirow{2}{*}{Architecture} & Attention & 32.9  & 3.9e-1  & 0.18  \\ 
                    & Adaptive AF   & 18.7 & 7.8e+0 & 0.22  \\ 
                    & FFT           & 19.8  & 1.6e+2 & 0.20  \\ \hline
                             
    \end{tabular}
\end{center}
\end{table}

% table with a comparison of the accuracy for different values.

\subsection{Two-Phase 1D Flooding at Constant Pressure Drop}

This section addresses the performance of PINNs in solving the problem of two-phase flow in porous media with constant pressure drop (eq. \ref{eq:systemof1d2pflow}), which requires a coupled solution of PDEs governing both phases. The complexity of this problem lies in accurately capturing the dynamics of shock fronts and breakthrough behavior in the flow. The loss landscape of this problem is expected to be more complex than that of the Buckley-Leverett problem, as it involves solving two distinct PDEs simultaneously, each governed by two state variables. We employed the same system properties as described in the previous section (Table \ref{table:systemprops}), except that instead of constant injection rate, constant inlet/outlet pressures of 400/200 psi were used. Also, distinct neural networks were utilized to model the state variables of the system, specifically for water saturation ($s_w$) and non-wetting phase pressure ($p_{nw}$). For the cases other than Vanilla-PINNs, the model were changed accordingly.

\begin{figure}[h!]
    \centering
    % \includesvg[width=6.5cm]{figs/base1Denplot.svg}
    \includegraphics[width=7cm]{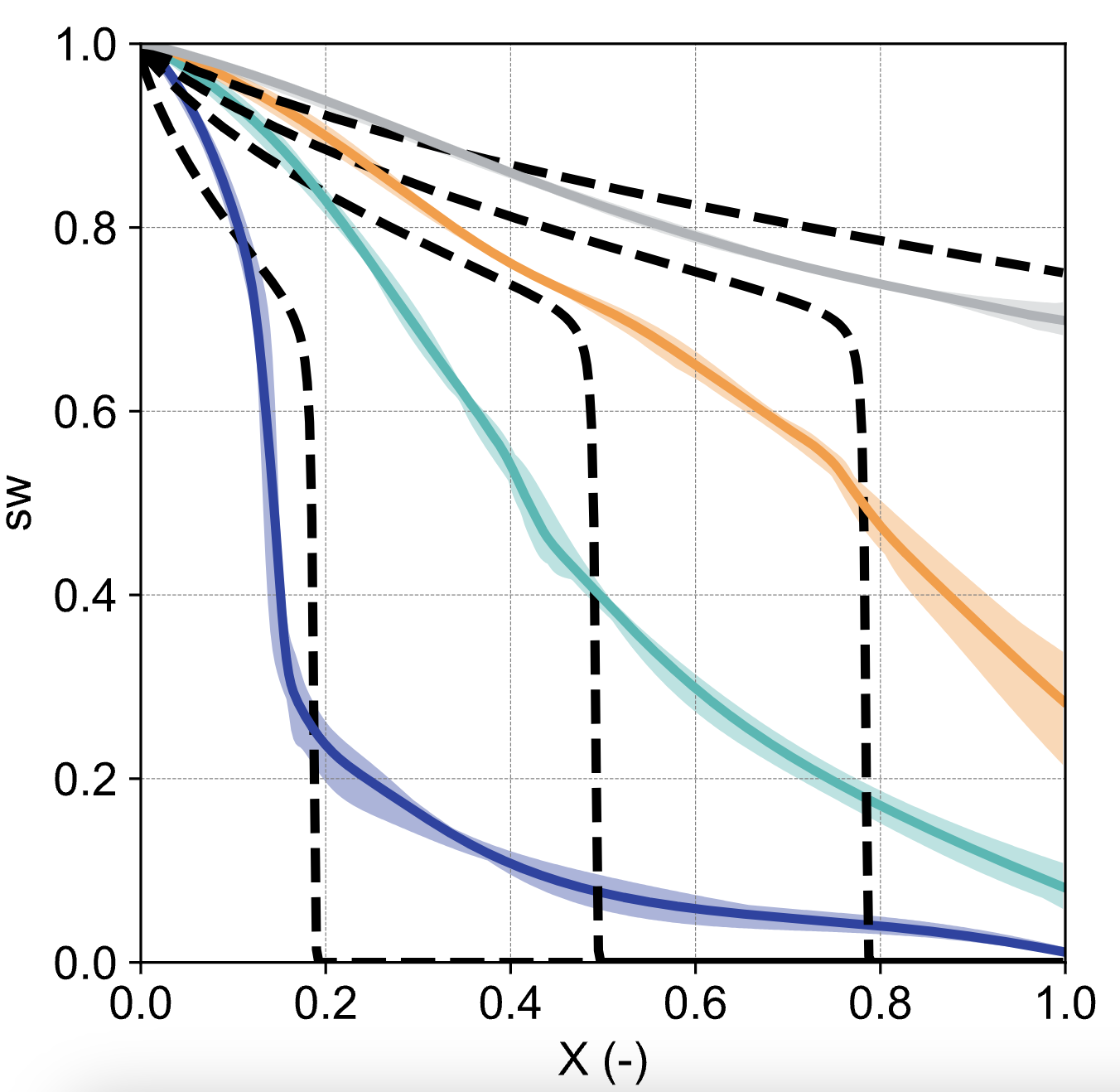}
    \caption{The accuracy of vanilla-PINNs in solving the coupled problem of 1D two-phase flow in porous media, in constant pressure boundary conditions. The solution is repeated for an ensemble of random initializations.}
    \label{fig:onedvanilla}
\end{figure}

\begin{figure}[h!]
    \centering
    \includegraphics[width=0.85\linewidth]{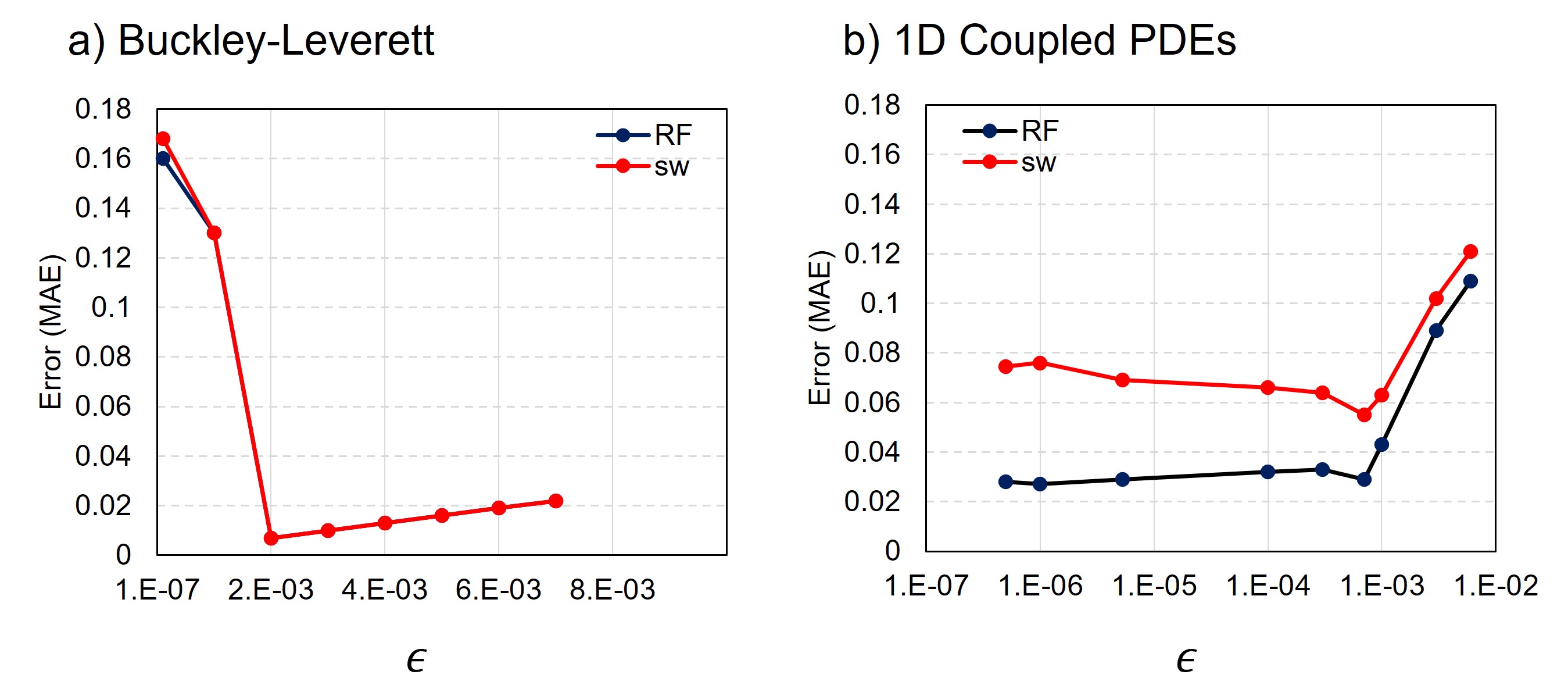}
    \caption{A sensitivity analysis on the impact of applying artificial viscosity with different $\epsilon_i$ values on the accuracy of the obtained solutions, for both $s_w$ and RF parameters. a) Buckley-Leverett, b) 1D Coupled PDEs}
    \label{fig:artificialviscepsilon}
\end{figure}

As visualized in Fig.~\ref{fig:onedvanilla}, the application of vanilla-PINNs to this problem yields suboptimal results, where PINNs fail to capture the expected shock fronts, a crucial aspect of two-phase flow dynamics. The PINNs solution smears out the sharp gradients at the fronts, leading to significant errors in the prediction and an early prediction of water breakthrough (it can also be seen in the calculated TV, in Fig. \ref{fig:tvdtrack}b, in which the TV value started to reduce quickly). However, after the breakthrough point — where the shock front disappears — the local error in the solution becomes acceptable, indicating that vanilla-PINNs can perform reasonably well in post-shock conditions. To address the limitations of vanilla-PINNs in capturing shock fronts, several improvements were tested. Table~\ref{table:1D2P_compare} presents a benchmark comparison of the accuracy of solutions achieved by these techniques. Each method is evaluated based on its ability to resolve the shock front and its overall performance in solving the 1D two-phase flow problem. The results indicate that the applied techniques struggle to effectively capture the shock front with acceptable accuracy. However, certain approaches, such as the addition of an artificial viscosity term (see Fig. \ref{fig:artificialviscepsilon} for the sensitivity to the coefficient values) or the use of the WENO (Weighted Essentially Non-Oscillatory) technique, demonstrate a slight reduction in the error levels for the calculated water saturation ($s_w$), which is the parameter of primary interest. These methods show promise in improving the solution accuracy, particularly in mitigating errors associated with shock front resolution.

\begin{table}[ht]

\begin{center}
\caption{A comparison of the performance of different methods for forward simulation of 1D two-phase flow PDE.}
    \label{table:1D2P_compare}
    \begin{tabular}[t]{llcccc}
    \hline
    \textbf{Group} & \textbf{Method} &  \makecell[c]{\textbf{Run-time} \\ (ms/ep.)} & \makecell[c]{\textbf{$L_t$ } \\ (MAE)} & \makecell[c]{\textbf{RF} \\ (MAE)} & \makecell[c]{\textbf{$s_w$ } \\ (MAE)}  \\ \hline
    
    Vanilla PINNs & - & 77.25 & 3.0e+01 & 0.027 & 0.080 \\ \hline
    \multirow{2}{*}{Physics-Based} 
                    & Artificial Viscosity & 94.4 & 5.1e+01 & 0.029 &  0.055 \\ 
                    & Entropy & 78.4 & 3.8e+01 & 0.029 & 0.072 \\ \hline
    
    \multirow{1}{*}{Training} 
                    & WENO & 79.1 & 5.2e+01 & 0.030 & 0.068 \\     
                    & RAR-D &  90.2 &  6.0e+01  &  0.027 & 0.079  \\ \hline
                    
    \multirow{2}{*}{Architecture} 

                    & Attention & 89.1 & 2.8e+01 & 0.020 & 0.064 \\ 
                    & Relaxation & 61.4  & 6.88e+01 & 0.017 & 0.093 \\ 
                    & Adaptive AF & 89.6 & 2.97e+01 & 0.020 & 0.074 \\ 
                    & FFT &  86.1 & 2.45e+01 & 0.028 & 0.061 \\ \hline
    \end{tabular}
\end{center}
\end{table}

\section{Discussion and Conclusions}

In this review, we assessed the capabilities and limitations of Physics-Informed Neural Networks (PINNs) in modeling partial differential equations (PDEs) with discontinuities, such as shock waves, with a particular emphasis on multiphase flow in porous media. Our analysis included a benchmark study of applying these methods to solve two-phase flow problems in porous media, encompassing two levels of complexity in the underlying PDEs. The case studies confirmed that Vanilla-PINNs fail to capture shock effects effectively, leading to significant mass conservation errors. By reviewing the literature, we categorized shock management techniques into three primary groups: \textbf{1)} physics-based modifications, \textbf{2)} loss function and training modifications, and \textbf{3)} architecture modifications. The benchmarking study demonstrated varying levels of effectiveness among these methods in addressing discontinuities, highlighting their respective strengths and limitations.

Physics-based techniques, such as incorporating artificial viscosity terms or using weak formulations (e.g., wPINNs), modify the governing PDEs to reduce their susceptibility to shock-related issues (infinite gradients). These methods smooth discontinuities and mitigate the effects of infinite gradients, improving numerical stability and accuracy. However, they come with increased computational costs, particularly in high-dimensional systems, and their effectiveness in complex, real-world problems remains unproven. While these techniques have shown promise in simpler PDEs like the Buckley-Leverett (BL) equation, they struggle to handle fully coupled two-phase flow equations effectively.

Loss function and training modifications focus on enhancing the optimization process to better handle shocks and discontinuities. Techniques such as residual weighting, gradient annihilation, and adaptive collocation point sampling redistribute computational effort to more/less critical regions, reducing errors near shocks. However, these methods require careful hyperparameter tuning and often struggle with accurate shock positioning in complex scenarios. Despite these challenges, they are computationally efficient and relatively straightforward to implement, making them suitable for a variety of problems.

Architecture-based techniques, such as domain decomposition (e.g., XPINNs), attention mechanisms, and Fourier feature embeddings, aim to improve the neural network's ability to handle high gradients and quasi-discontinuities. While these methods enhance flexibility and mitigate issues like spectral bias, they fail to fully resolve challenges such as infinite gradients. Additionally, the added computational overhead limits their scalability to high-dimensional problems. Domain decomposition approaches are effective for static discontinuities but face significant challenges when applied to spatiotemporal dynamic discontinuities.

Our review highlights several key points:
\begin{itemize}
    \item \textbf{Vanilla PINNs are inadequate for handling discontinuities}, but significant progress has been made through physics-based, loss function, and architecture-based modifications. These advancements have enabled PINNs to better capture sharp features and reduce errors.
    
    \item \textbf{Physics-based modifications} improve stability and accuracy but come with increased computational costs and limited applicability to complex systems.
    
    \item \textbf{Loss function modifications} enhance optimization but require careful tuning and struggle with shock positioning in complex problems.
    
    \item \textbf{Architecture-based techniques} improve the network's ability to handle high gradients but introduce computational overhead and fail to fully resolve infinite gradients.
    
    \item \textbf{The curse of dimensionality significantly limits the performance of current shock management techniques in PINNs.} While current methods are effective for simplified or small-scale problems, their applicability to complex, high-dimensional systems remains unproven. Challenges persist in accurately capturing shock fronts, ensuring mass conservation, and scaling to real-world scenarios.
\end{itemize}

At the end, like traditional numerical methods, PINNs face challenges when dealing with discontinuities in PDEs. While significant improvements have been made in numerical methods over the past few decades, PINNs are still in their early stages of development. Many techniques used for shock modeling in PINNs, such as artificial viscosity or WENO methods, were originally developed for traditional numerical methods. However, the nature of errors around shock fronts differs between the two: PINNs tend to smear out shock fronts, whereas numerical methods often exhibit oscillations near these fronts. Recognizing these differences is crucial for fostering knowledge exchange between the two fields. Moving forward, insights and advancements from either family of methods can contribute to the refinement of shock management techniques, ultimately enhancing the robustness and accuracy of both approaches.

In conclusion, while significant progress has been made in improving PINNs' ability to handle discontinuities, challenges remain in scaling these methods to complex, high-dimensional systems. Future research should focus on addressing the curse of dimensionality, improving computational efficiency, and validating these techniques in real-world applications. By bridging the gap between traditional numerical methods and PINNs, we can unlock their full potential for solving complex PDEs with discontinuities.

\subsection{Future Directions}

Looking ahead, future research could benefit from a deeper exploration of hybrid methodologies that combine multiple techniques—for example, integrating weak formulations with adaptive architectures or residual-based adaptive sampling—to further enhance PINNs' stability and precision in addressing complex shock dynamics. Additionally, embedding domain-specific constraints and advancing real-time adaptive strategies may significantly extend PINNs' applicability to high-dimensional, dynamic field scenarios. Automated parameter tuning also holds promise for reducing computational overhead, accelerating the deployment of PINNs in industrial simulations and decision-making processes. In particular, moving to alternative methods based on combination with numerical techniques, such as mixed PINNs \cite{Harandi2023MixedDomains}, or using discrete loss terms \cite{Saqlain2023DiscoveringPINNs, Karnakov2022OptimizingTools}, could allow the direct integration of strategies proven effective in numerical simulations, providing a more robust approach to PINNs for handling shock front complexities. In addition, a greater focus on physics-informed
operator learning models \cite{DeRyck2024NumericalLearning} may be a suitable approach to provide more reliable solutions. Furthermore, future research efforts should prioritise expanding the focus to higher-dimensional problems characterised by more complex loss landscapes, as there remains a significant research gap in addressing these types of challenges.

\section*{Acknowledgments}
Abbasi, Andersen, and Hiorth acknowledge the Research Council of Norway and the industry partners of NCS2030 (RCN project number 331644) for their support. Abbasi and Andersen extend their gratitude to the Research Council of Norway for funding the "Pisces-AI: Physics-Informed AI for Subsurface Characterization Experiments" project (RCN Project number 354776).

\bibliography{DLREVIEW_CITEDRIVE,PAL_ANDERSEN,Mendeley_Jassem} 
% \input{main.bbl}
% \bibliography{main.bbl}
% \bibliographystyle{unsrt}

% \section*{Applied Meta-parameters}

% \subsection*{BL}

% \begin{itemize}
%     \item \textbf{WENO}: $\omega=0.45$
% \end{itemize}

% ...

% \section*{Appendix 3: Computational tools}

% \section*{Metrics}

% \subparagraph{Uncertainty Quantification}
% The understanding and characterisation of uncertainties in models are of interest in problems related to flow in porous media due to the natural uncertainties inherent in the involved processes. Due to the statistical nature of physics-based deep learning methods, different approaches for uncertainty quantification problems have been introduced in recent years, such as ensemble-based  \cite{Aliakbari2023EnsembleDomains}, and Bayesian approaches \cite{Yang2021B-PINNs:Data}. Some works tried to take advantage of these approaches for the porous media applications. 

\appendix
\section{Two-Phase 1D Flooding in Porous Media at Constant Pressure Drop}
\label{app:constant_pressure}

To arrive at the classic problem describing flooding to displace a fluid from a 1D porous medium, we limit the number of phases to two: the invading phase, assumed to be wetting ($w$), and the invaded phase, assumed to be non-wetting ($nw$). We assume the domain to be homogeneous, with constant porosity ($\phi$) and permeability ($K$), and consider the fluids to be incompressible. Considering the system to be horizontal, gravity can be eliminated and finally we note that capillary pressure is set to zero on large scales (e.g. interwell distances of 100's toward 1000 m), such that $p_w = p_{nw}$. The mass conservation equations, after eliminating two variables with the constraints, simplify to:

\begin{equation}
\begin{cases}
    \phi  \frac{\partial s_w}{\partial t} - \frac{K }{\mu_w}  \frac{d}{dx} \left( k_{rw} \frac{d p_{nw}}{dx} \right) = 0, \\
    \phi \frac{\partial (1- s_{w})}{\partial t} - \frac{K }{\mu_{nw}} \frac{d}{dx} \left( k_{rnw}  \frac{d p_{nw}}{dx} \right) = 0, \\
\end{cases}
,\ \ \ x \in \Omega
\label{eq:systemof1d2pflow_appendix}
\end{equation}

These equations must be solved simultaneously as a coupled system to capture the inter-dependencies between the phases. The independent variables of the system are $x$, and $t$, and the dependent (state) variables are $s_w$ and $p_{nw}$. 
Assuming an initial condition (IC) where \( s_w(x, t=0) = s_{w,\text{initial}} \), we solve a water flooding case with boundary conditions (BC) of wetting phase injection at the inlet (\(x=0\)), free production at the outlet (\(x=L\)) driven by a constant pressure difference between the boundaries.

\begin{equation}
p (x=0,t) = p_{\text{inlet}}, \quad s_w (x=0,t) = s_{w,\text{inlet}},
\end{equation}
\begin{equation}
p (x=L,t) = p_{\text{outlet}}, 
\end{equation}

With capillary and gravitational forces ignored, the result represents viscous multiphase flow in porous media. The injection rate will vary with time according to how the mobility in the system changes. Here, the wetting phase displaces the non-wetting phase, forming a sharp interface between the two phases as part of the solution. This interface appears as a shock wave, commonly known as a shock front. In the above equations, the relative permeability function can be determined using the below equation,
\begin{equation}
k_{ri} = k_{ri}^{max}S_i^{n_i}
\label{eq:relperm}
\end{equation}
where $k_{ri}^{max}$ is the phase maximum relative permeability and $n_i$ is the saturation exponent. Also, $S_i$ is the mobile phase saturation,
\begin{align}
{S}_{i} = \frac{{s}_{i} - {s}_{wc}}{1 - {s}_{nwr} - {s}_{wc}};
\label{eq:Sweq}
\end{align}
where ${s}_{wc}$ and ${s}_{nwr}$ are connate wetting phase saturation, and irreducible non-wetting phase saturation, respectively. 

\section{Two-Phase 1D Flooding in Porous Media at Constant Injection Rate: The Buckley-Leverett Problem}
\label{app:buckley_leverett}

Based on the same assumptions as above, Buckley and Leverett (BL) \cite{EBuckley1942MechanismSands} combined the phase equations in eq. \ref{eq:systemof1d2pflow_appendix}, and eliminated the pressure variable by introducing the total Darcy velocity $v_t=v_w+v_{nw}$ (which can be shown to be uniform, $\partial_x v_t = 0$). At constant rate injection conditions $\partial_t v_t=0$ (instead of constant pressure drop), the number of state variables is reduced to one ($s_w$) and we can express the flow equation as:

\begin{equation}
\ensuremath{  \frac{\partial {s_{w}}}{\partial t} + \frac{v_t}{\phi}\frac{d {f_{w}}}{d s_w}\frac{\partial {s_{w}}}{\partial x} = 0 ,\ \ 
\left(i=w,nw\right) }
\label{eq:BLeq_appendix}
\end{equation}
where \(f_w(s_w)\) is the fractional flow function of the wetting phase, defined as
\begin{equation}
\fractionalflow
\end{equation}
The eq. \ref{eq:BLeq_appendix} is hyperbolic, producing wave-like solutions that include the formation of a shock front followed by a rarefaction wave behind the front. The shock extends between a saturation \( s_{ws} \) (the saturation at which the shock front appears) and the initial saturation \( s_{wi} \). The method of characteristics provides the position of each water saturation located behind the front:
\begin{align}
x = \frac{v_t}{\phi} \frac{d f_w}{d s_w} t
\label{eq:shocklocationBL}
\end{align}

The saturation profile $s_w(x)$ is identical for the two systems eq. \ref{eq:systemof1d2pflow_appendix} and eq. \ref{eq:BLeq_appendix} when the same amount of water has been injected. However, in the first system we need to solve for pressure and saturation simultaneously to determine the amount of water injected at a given time, while in the second system, the amount of water injected follows directly from the rate and time and the saturation profile can be solved independently.  

A semi-analytical graphical approach to solving eq. \ref{eq:BLeq} is the Welge tangent method \citep{Welge1952ADrive}, based on the Rankine-Hugoniot conditions. This method determines the shock front saturation by drawing a tangent from the initial water saturation point to the fractional flow curve, demonstrated in Fig. \ref{fig:welge}.  The saturation in front of the shock wave is equal to $s_{wi}$. To determine the shock saturation \( s_{ws} \), the conditions of mass balance and velocity continuity (between the front and continuous solution) result in the following equation:
\begin{align}
\left. \frac{d f_w}{d s_w} \right|_{s_w = s_{ws}} = \frac{f_w(s_{wf}) - f_w(s_{wi})}{s_{ws} - s_{wi}}
\end{align}

\end{document}